\documentclass[preprint,12pt]{elsarticle}

\usepackage{amssymb}
\usepackage{amsmath}
\usepackage{esvect}
\usepackage{url}
\usepackage{stfloats}
\usepackage{morefloats}
\usepackage{times} 
\usepackage{ifpdf}
\usepackage{graphicx}
\graphicspath{{Pictures/}{/}}

\usepackage{setspace}
\usepackage{appendix}
\usepackage{todonotes}
\usepackage{hyperref}
\usepackage{algorithmic}
\usepackage[linesnumbered,nofillcomment,commentsnumbered]{algorithm2e}
\usepackage{subcaption}

\usepackage{siunitx}
\usepackage{tikzpagenodes}
\usepackage{balance}

\usepackage{multirow} 

\newcommand{\revJAC}[1]{\textcolor{black}{#1}}
\newcommand{\revJACS}[1]{\textcolor{black}{#1}}

\newcommand{\method}{FS-Planner}

\journal{Robotics and Autonomous Systems}
\begin{document}

\begin{frontmatter}

\title{Exploiting Euclidean Distance Field Properties for Fast and Safe 3D planning with a modified Lazy Theta*}

\author{Jose A. Cobano, L. Merino and F. Caballero} 
\address{Service Robotics Lab (SRL) \\ University Pablo de Olavide \\ Seville, Spain}
\ead{jacobsua@upo.es, lmercab@upo.es, fcaballero@upo.es}


\begin{abstract}
\revJAC{This paper presents the FS-Planner, a fast graph-search planner based on a modified Lazy Theta* algorithm that exploits the analytical properties of Euclidean Distance Fields (EDFs). We introduce a new cost function that integrates an EDF-based term proven to satisfy the triangle inequality, enabling efficient parent selection and reducing computation time while generating safe paths} \revJACS{with smaller heading variations.} \revJAC{We also derive an analytic approximation of the EDF integral along a segment and analyze the influence of the line-of-sight limit on the approximation error, motivating the use of a bounded visibility range. Furthermore, we propose a gradient-based neighbour-selection mechanism that decreases the number of explored nodes and improves computational performance} \revJACS{without degrading safety or path quality.} \revJAC{The FS-Planner produces safe paths} \revJACS{with small heading changes} \revJAC{without requiring the use of post-processing methods. Extensive experiments and comparisons in challenging 3D indoor simulation environments}\revJACS{, complemented by tests in real-world outdoor environments,} \revJAC{are used to evaluate and validate the {\method}. The results show consistent improvements in computation time, exploration efficiency, safety, and smoothness} \revJACS{in a geometric sense} \revJAC{compared with baseline heuristic planners, while maintaining sub-optimality within acceptable bounds. Finally, the proposed EDF-based cost formulation is orthogonal to the underlying search method and can be incorporated into other planning paradigms.}
\end{abstract}


\end{frontmatter}



\section{Introduction}
\label{sec:intro}


\subsection{Motivation}

\begin{figure}[!t]
	\centering
    \includegraphics[width=.99\textwidth]{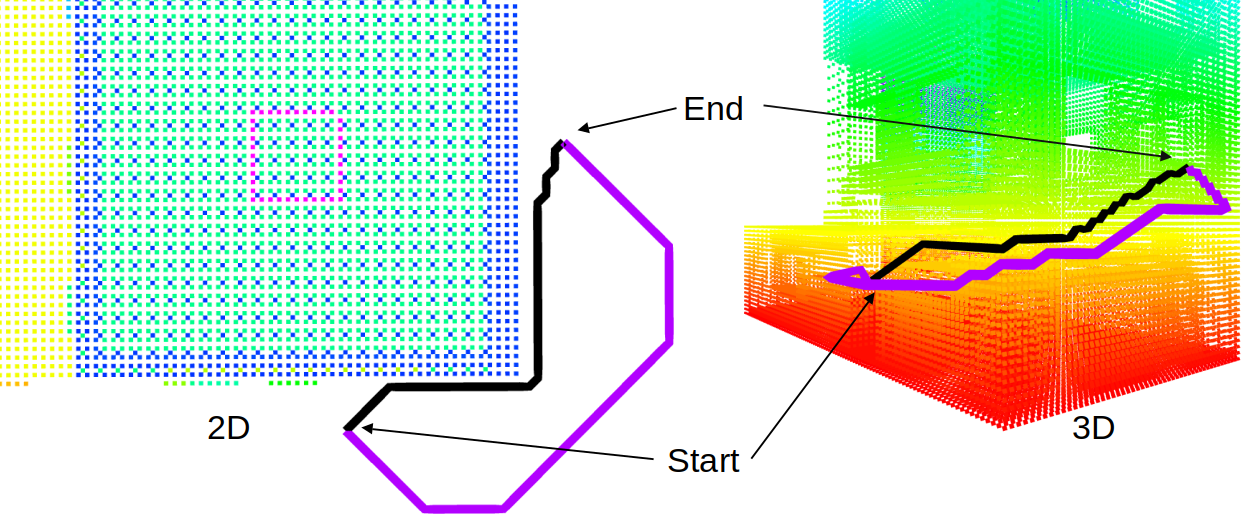}
	\caption{Paths computed by original A* algorithm in \revJAC{black}, and computed by A*+M1 from \cite{iros2022_edf} in \revJAC{purple} considering distance to obstacles represented as EDF: 2D (left) and 3D (right). Both paths show how the distance to obstacles influences on the computed path by A*+M1.}
	\label{fig:motivation}
\end{figure}

Path planning in complex 3D environments is a basic procedure that robots must carry out frequently in numerous nowadays applications: tunnel inspection \cite{Elmokadem21}, urban firefighting \cite{fr21mbzirc}, subterranean inspection, like the DARPA Sub-T challenge\footnote{ \href{https://www.darpa.mil/program/darpa-subterranean-challenge}{https://www.darpa.mil/program/darpa-subterranean-challenge}} 
or drone racing challenges such as AlphaPilot \cite{Foehn2022}. Approaches ranging from sampling-based to searching-based planners have been successfully applied in the state of the art. We can also find methods based on curve interpolation and optimization. Graph search and heuristic planners such as A* \cite{Ha_2021} or Theta* \cite{Choi2010} are widely used. These methods usually focus on the generation of optimal paths with respect to the path length, leaving aside safety and 
smoothness of the path (see Fig. \ref{fig:motivation}). As a consequence, computed paths tend to include unnecessary heading changes, which requires post-processing methods \cite{Ferguson2006UsingIT,Li_2023_ral} to mitigate such effect. 

Euclidean Distance Fields (EDF) are a popular environment representation. The field gives the distance to the closest obstacle for any point in space. EDFs have interesting analytical properties 
\cite{Jones:TVCG2006}. They are particularly useful in 3D path planning \cite{Oleynikova2016,Gao2017,Usenko2017,Zhu2021}, but to the best for our knowledge, most existing path planning approaches that integrate EDFs do not exploit their analytical properties. 
Despite the usefulness of EDF in path planning, heuristic search planners are rarely integrated with them to leverage their properties \cite{iros2022_edf}. Fig. \ref{fig:motivation} shows the different paths computed by the A* algorithm using a grid map as obstacle representation and the path length as cost, and the A*+M1 \cite{iros2022_edf} which considers an EDF as environment representation and a cost function that integrates costs related to the distance to obstacles. Notice how considering distance to obstacles using EDF as representation enables safer \revJACS{and geometrically smoother paths (in terms of heading changes)} \cite{iros2022_edf}. 
On the other hand, the path length as cost holds the triangle inequality, which states that the sum of the lengths of two sides of a triangle must be longer than the length of the remaining side. It can be used in heuristic planners to simplify the calculations to select the parent node. 


The goal of this paper is to develop efficient and safe heuristic path planners able to integrate an EDF, consider an EDF-based cost function 
and take advantage of its properties in order to directly compute safe and \revJACS{geometrically} smooth paths. 
It will be also shown that we can even modify existing heuristic planners as both A* and Lazy Theta* planning algorithms to naturally exploit the EDF properties. This will enable approximation of the EDF cost along a segment\revJAC{, in addition to using the EDF gradient to identify directions that naturally lead the planner away from obstacles, thereby reducing} the number of visibility neighbours \revJAC{and the corresponding} computation time. Furthermore, \revJAC{this reduction in neighbour exploration can improve path smoothness} \revJACS{with fewer abrupt heading changes}\revJAC{, decreasing the need for any }post-processing methods. Finally, if we can demonstrate that the proposed EDF-based cost function holds the triangle inequality\revJAC{, which preserves the parent-selection efficiency of Lazy Theta*}, the simplification of the calculations will also be ensured. This would be an extension of the previous work \cite{iros2022_edf} because its cost function does not satisfy the triangle inequality \revJAC{and therefore requires additional parent comparisons}.

\subsection{Related work}

Among the searching-based methods, A* stands as one of the most used solutions. Theta* is a variant of A* that shortens the path length \cite{Daniel2010}, and Lazy-Theta* improves its efficiency \cite{THETA}. 
The authors in \cite{Choi2010} present a 2D Theta* planner that deal with non-uniform costmaps in 2D, but the paths are not smooth and the triangle inequality is not held, so they have to generalize the original Theta* algorithm and more calculations to select the parent node are performed. Moreover, unnecessary heading changes still occur. In \cite{Li_2023_ral} the smoothness of the path is improved with an A* algorithm but they need a post-processing algorithm.

There are also approaches that use traversal cost with 
Rapidly-exploring Random Trees (RRT)\cite{Jaillet_2010} but they also need further post-processing to be smooth. Other approaches compute a path based on heuristic search and linear quadratic minimum-time control but then the initial path has to be refined in a carefully designed B-spline optimization to achieve smooth and safe paths \cite{Zhou_Gao2019}. In other cases, once the initial path is generated, gradient-based methods like CHOMP selects a smoother and safer path \cite{Ratliff2009}. 

Oleynikova et al. \cite{Oleynikova2016} use Euclidean Signed Distance Field (ESDF) to consider the distance cost to the obstacles using Informed RRT*, but the planned trajectory can vary significantly from execution to execution. \revJAC{On the other hand, the proposed path planner is deterministic, producing the same solution under identical conditions. Moreover, \cite{Oleynikova2016} uses the EDF solely as a collision-avoidance metric, while the proposed path planner integrates an EDF-based cost function and exploits the analytical properties of the EDF for both cost evaluation and neighbour selection.} 
The authors in \cite{Gao2018} adopt a fast marching path planning method on velocity fields using ESDF, but they use Bernstein basis polynomials to generate smooth paths. Moreover, fast marching methods tend to ﬁnd longer paths than searching-based algorithms as A* since the metric is velocity in the field, not the path length\revJAC{, whereas the proposed path planner directly balances distance, safety, and} \revJACS{geometric} \revJAC{smoothness. Furthermore, \cite{Gao2018} requires a post-processing smoothing stage, while the proposed path planner can generate smooth paths directly. Finally, fast marching methods scale poorly in cluttered 3D environments}. EDFs are also integrated as environment representation in heuristic planners in the previous work of the authors \cite{iros2022_edf} so that safe paths \revJACS{with fewer abrupt heading changes} are directly generated with a Lazy Theta* algorithm. \revJAC{The advantages of the proposed path planner with respect to \cite{iros2022_edf} are highlighted in Section \ref{contributions}.}


\subsection{Contributions}
\label{contributions}

The main contributions are:
\begin{itemize}
    \item A new fast and cost-aware heuristic planner based on a modified Lazy Theta* planning algorithm that integrates an EDF previously computed as environment representation and exploits its properties. 
    \item A new cost function considering the distance to the obstacles so that the triangle inequality is held.
    \item The computation time of the new EDF-based distance cost term is considerably reduced by leveraging the EDF properties.
    \item A modified Lazy Theta* algorithm with a novel exploration of the visibility neighbours that foster velocity and \revJAC{improve path} \revJACS{smoothness in a geometric sense,} \revJAC{which} reduces the need for post-processing methods.
\end{itemize}

This paper extends the authors' previous work \cite{iros2022_edf} by further leveraging the EDF properties, which lead to modifications in the Lazy Theta* algorithm. \revJAC{First, a distance EDF-based cost function} is introduced \revJAC{and satisfies the triangle inequality; in contrast, \cite{iros2022_edf} does not satisfy this property. Second, the proposed path planner reduces the visibility-neighbour set using the EDF gradient, which substantially decreases the number of explored nodes and leads to faster computation,} producing \revJACS{geometrically} smoother and safer paths than those obtained with \cite{iros2022_edf}. 
How visibility neighbours are selected also affects safety and smoothness. \revJAC{Additionally, unlike \cite{iros2022_edf}, the proposed path planner provides a more detailed theoretical justification of the EDF properties, explicitly covering the case of non-convex obstacles through convex decomposition (see Section \ref{computation_O}).}

The planner is 
evaluated with many random tests in challenging 3D environments. 
The implementation of the algorithm and the scripts to launch the tests are open sourced\footnote{ \href{https://github.com/robotics-upo/Heuristic\_path\_planners/}{https://github.com/robotics-upo/Heuristic\_path\_planners/}}.

The paper is organized into five sections. Section II describes the proposed 3D planner. The new cost function and the EDF properties considered are presented in Section III. The selection of the visibility neighbours is described in Section IV. The results and analysis 
are shown in Section V. Finally, Section VI presents the conclusions.

\section{Fast and Safe Path Planner Overview} \label{planners} 

This section describes the proposed fast and safe path planner (FS-Planner). It is based on the Lazy Theta* algorithm \cite{THETA}, and it leverages the properties of the EDF. Lazy Theta* is a variation of Theta*, which is, in itself, an any-angle algorithm based on A* \cite{Hart1968}. Theta* interleaves search and smoothing by including into A* visibility checks that can find new parent nodes if there is a line of sight \cite{Daniel2010}.  In Theta*, the line of sight checks increase the computational load, thus negatively impacting the  performance in terms of computation time. The original Lazy Theta* reduces the line of sight checks performed during the search to those strictly necessary. It assumes line of sight between nodes when expanding the search and the triangle inequality is held by the cost function, the path length in this case, 
delaying the line of sight check. 
This is relevant in 3D due to the number of visibility neighbours for each node.

The proposed planner integrates an EDF to efficiently represent the robot's environment and to define the graph employed in the planner. The 3D EDF voxel grid is generated offline considering the distance to the closest \revJAC{obstacle} for every node in a map with fixed resolution. 
As we will see, due to the EDF properties, we can obtain an approximation of the EDF cost along segments (see Section \ref{sec:cost_function}) and the EDF representation enables easy obstacle gradient computation based on local EDF derivative (see Section \ref{selection}). 

Algorithm \ref{lazy_proposed} shows the modifications performed in the original Lazy Theta* algorithm (highlighted in red). First of all, a cost function that takes into account the distance to obstacles from the EDF (see $ComputeCost$ function of Algorithm \ref{lazy_proposed}) is used with the aim of computing safer paths (the cost function is described in Section \ref{sec:cost_function}). 
Another necessary modification is to limit the line of sight ($LoS$) in the $SetVertex$ function of Algorithm \ref{lazy_proposed} in order to generate \revJACS{geometrically} smoother paths and foster safety. 
Otherwise, different nodes of the path could be linked if they are visible to each other, overriding the possible benefits of introducing the distance cost term in order to generate safer paths. 
These modifications imply an extension of the algorithm rather than a modification on how it works. 

\begin{algorithm}
 \small
 \SetKwProg{UpdateVertex}{UpdateVertex(s,s')}{}{end}
 \SetKwProg{ComputeCost}{ComputeCost(s,s')}{}{end}
 \SetKwProg{SetVertex}{SetVertex(s)}{}{end}
 \SetKwProg{IF}{If}{ then}{}
 \SetKwProg{FOREACH}{foreach}{ do}{}
 \SetKwProg{WHILE}{while}{ do}{}
 \SetKwProg{main}{Main()}{}{end}
  \main{}{
    open := closed := $\emptyset$\;
    g($s_{start}$) := 0\;
    parent($s_{start}$):=$s_{start}$\;
    open.Insert($s_{start}$,$g(s_{start})+h(s_{start})$)\;
    \WHILE{open $\neq \emptyset$}{
    s := open.Pop()\;
    SetVertex(s); \tcc{Only Lazy Theta*} \
        \IF{ $s = s_{goal}$ }{
            \Return{"path found"}\;
        }
        closed := closed $\cup s$\;
        \FOREACH{$s' \in nghbr_{vis}(s)$ }{
            \tcc{Minimum EDF'} \ 
            \textcolor{red}{$s_{edf}$=ComputeMinDerivEDF($s$, $s'$)}\; 
        }            
        \tcc{Direction to choose the neighbours} \ 
        \textcolor{red}{$\hat{u}$=ComputeVector($\hat{s}_{edf}$,$\hat{s}_{goal}$)}\;
        \textcolor{red}{$nghbr_{edf}(s)$=ChooseNeighbours($\hat{u}$)}\;
        \FOREACH{\textcolor{red}{$s' \in nghbr_{edf}(s)$}}{
            \IF{$s' \not\in$ closed }{
                \IF{$s' \not\in$ open }{
                g($s'$) :=  $\infty$\;
                parent($s'$) := NULL\;
            }
            UpdateVertex($s$,$s'$)\;
            }
        }
     }
     \Return{"no path found"}\;
  }
  
  \UpdateVertex{}{
  $g_{old}$ := $g(s')$\;
  ComputeCost(s,s')\;
  
  \IF{$g(s') < g_{old}$}{ 
    \IF{$s' \in $ open}{ 
        open.Remove(s')\;
      }
      open.Insert($s'$,$g(s')+h(s')$)\;
    }
  }
  \ComputeCost{} {
    \tcc{Path 2}
    \IF{$g(parent(s)) + ||s'-parent(s)||+ \textcolor{red}{\frac{c_w}{O(parent(s),s')}} < g(s')$ }{
        $parent(s')$ := $parent(s)$\;
        $g(s')$ := $g(parent(s)) + ||s'-parent(s)||+ \textcolor{red}{\frac{c_w}{O(parent(s),s')}}$ \;
        }
  }
  \SetVertex{}{
    \IF{NOT lineofsight(parent(s),s) \textcolor{red}{{OR lineofsight(parent(s),s) $>$ max}}}{
        \tcc{Path 1}
        $parent(s)$ := $\arg \min_{s' \in nghbr_{vis} \cap closed} (g(s')+||s'-s||+ \textcolor{red}{\frac{c_w}{O(parent(s),s')}})$\;
        $g(s)$ := $\min_{s' \in nghbr_{vis} \cap closed}(g(s')+||s'-s||+ \textcolor{red}{\frac{c_w}{O(parent(s),s')}})$\;
        }
    }
\caption{FS-Planner proposed based on modified Lazy Theta*.}
\label{lazy_proposed}
\end{algorithm}

The second modification aims to reduce the visibility neighbours to explore around the current node by means of the obstacle-gradient information (see lines 12-17 in Algorithm \ref{lazy_proposed}). Depending on how the visibility neighbours are chosen, fast and safe paths can be generated while we explore fewer neighbours. 
This change does modify the algorithm behavior because the exploration is not performed on all the visibility neighbours as in the original Lazy Theta* algorithm. The selection of the neighbours to explore is explained in Section \ref{selection}. 

\section{The cost function} \label{sec:cost_function} 

This section presents the new cost function, and how its computation exploits the EDF's properties. 
The accumulated cost of a given node in the path in heuristic planners is given by:
\begin{equation}
\label{eq:cost_function}
    g(s_{i+1})=g(s_{i})+c(s_i,s_{i+1})
\end{equation}

\noindent where $g(s_{i})$ is the cost of the parent node $s_i$ and $c(s_i,s_{i+1})$ is the cost associated with moving from $s_{i}$ to $s_{i+1}$. 

Depending on how you choose $c(s_i,s_{i+1})$, you can penalize or favour different paths considering EDFs. 
In our approach, we define it as follows ($ComputeCosts$ in line 34 of Algorithm \ref{lazy_proposed} and $SetVertex$ in line 40 of Algorithm \ref{lazy_proposed}):
\begin{equation}
\label{eq:new_cost}
    c(s_i,s_{i+1})=||s_{i+1}-s_{i}||+\frac{c_w}{O(s_{i},s_{i+1})}
\end{equation}

\noindent where $||s_{i+1}-s_{i}||$ is the 
Euclidean distance between $s_{i}$ and $s_{i+1}$ (the usual term considered in the original Lazy Theta*), $c_w$ is a cost weight factor and $O(s_{i},s_{i+1})$ is the integral of the EDF along the segment defined by $s_{i}$ and $s_{i+1}$. 

Notice how the EDF of a node is greater the farther it is from an obstacle; thus, paths are safer when the integral term $O(s_{i},s_{i+1})$ is greater. Considering that the planner minimizes the cost $g$, we inverted the integral of EDF into the cost function as $O(s_{i},s_{i+1})^{-1}$.

Next sections detail the computation of the integral cost over the EDF, and demonstrate that the proposed cost function complies with the triangular inequality. 

\subsection{Computation of the integral of the EDF along the segment}
\label{computation_O}

The component of the cost containing $O(s_{i},s'_{i+1})$ depends on the EDF. The EDF is computed from the distance function to the obstacles, which is defined as: 

\textbf{Definition} Let $\Omega \subset \mathbb R^n$ be a nonempty set. Given any point $x \in \mathbb R^n$, then the distance from $x$ to the set $\Omega$ is given by the function:
\begin{equation}
    \label{eq_dist_func}
    d(x;\Omega)=inf\{\|x-y\| \; : \; y \in \Omega\}
\end{equation}

\noindent An important property of the distance function is that it is Lipschitz continuous with Lipschitz constant equal to 1 \cite[Proposition 1.74]{Mordukhovich2013AnEP}, and its gradient norm is $\|\nabla_x d(x;\Omega)\|=1$.

\begin{figure}[!t]
	\centering
    \includegraphics[width=.4\textwidth]{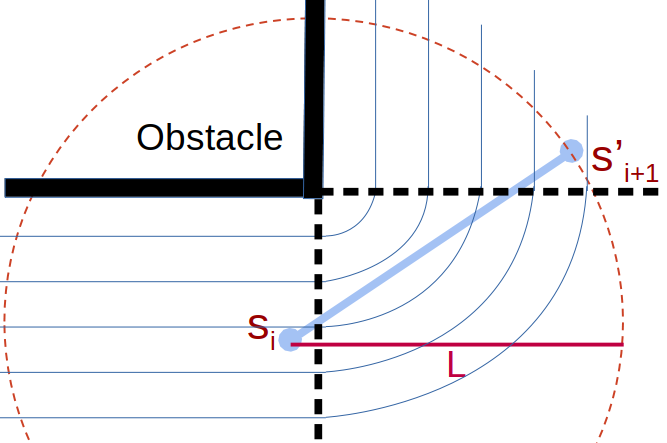}
	\caption{Computation of EDF integration for segments of length $L$ with a convex obstacle. The isocurves of the distance to obstacle are shown in blue.}
	\label{fig:point_convex}
\end{figure}

The cost $O(s_{i},s'_{i+1})$ is defined as the integral of the distance function along the straight segment $l$ that joins $s_{i}$ and $s'_{i+1}$ (see Fig. \ref{fig:point_convex}):

\begin{equation}
\label{eq:cost}
    O(s_{i},s'_{i+1})=\oint_l d(l;\Omega) dt=\int_{t_0}^{t_1}d(l(t);\Omega)\| l'(t)\| dt
\end{equation}

\noindent where $l(t)$ is the parameterized straight segment between $s_i=(x_i,y_i,z_i)$ and $s'_{i+1}=(x_{i+1}, y_{i+1}, z_{i+1})$, defined as:

\begin{eqnarray}
\label{eq:annex_segments1}
    l(t)=(x_i+(x_{i+1}-x_i)t,y_i+(y_{i+1}-y_i)t,z_i+(z_{i+1}-z_i)t)^T,\quad 0\leq t\leq 1 
\end{eqnarray}

\noindent so that $l(0)=s_i$ and $l(1)=s_{i+1}$. $l'(t)$ is the derivative of $l(t)$ with respect to the parameter, which can be easily seen that is $l'(t)=s_{i+1} - s_i$. Therefore,

\begin{equation}
    O(s_{i},s'_{i+1})=||s_{i+1}-s_{i}||\int_{0}^{1}d(l(t);\Omega)dt
\end{equation}

The computation of  $O(s_{i},s'_{i+1})$ has to be performed repeatedly in Algorithm 1 and it is a computationally expensive step. In the following we derive an approximation to this computation. 

Let's first assume that the obstacles defining the distance field form a convex set. In that case, we have the following:

\textbf{Proposition} The value of $O(s_{i},s'_{i+1})$, defined as (\ref{eq:cost}), in the presence of an obstacle represented by a convex set $\Omega$, is bounded as:

\begin{multline}
   \frac{d(s_i;\Omega)+d(s_{i+1};\Omega)}{2} ||s_{i+1}-s_{i}|| - \frac{||s_{i+1}-s_{i}||^{2}}{2} \leq O(s_{i},s'_{i+1})\leq \\ \leq \frac{d(s_i;\Omega)+d(s_{i+1};\Omega)}{2} ||s_{i+1}-s_{i}||
    \label{eq:bound}
\end{multline}

\textbf{Proof}: First of all, the distance function in $\mathbb R^n$ with respect to a convex set is a convex function (see \cite[Proposition 1.77]{Mordukhovich2013AnEP}). Therefore, $d(l(t);\Omega)$ is a convex function. The Hermite-Hadamard inequality \cite{dragomir2015inequalities} states that if a function $f: [a,b] \rightarrow \mathbb R$ is convex the following inequality holds:

\begin{equation}
    \label{eq:hermite-hadamard}
    (b-a) f(\frac{a+b}{2} ) \leq \int_{a}^{b} f(x)dx \leq (b-a) \frac{f(a)+f(b)}{2}
\end{equation}

\noindent The upper bound of this inequality directly leads to the upper bound in (\ref{eq:bound}).

Regarding the lower bound, we employ the Lipschitz property of the distance function, which states that the distance function at two given points cannot change faster than the distance between the two points. Thus, we have that for the distance function at the mid-point of the segment:

\begin{eqnarray}
    f(\frac{a+b}{2} ) = d(l(0.5);\Omega) \geq d(s_i;\Omega) - \frac{\|s_{i+1}-s_{i}\|}{2}\\
    f(\frac{a+b}{2} ) = d(l(0.5);\Omega) \geq d(s_{i+1};\Omega) - \frac{\|s_{i+1}-s_{i}\|}{2}
\end{eqnarray}

\noindent and by summing both equations we arrive to the lower bound in (\ref{eq:bound}).





Given the bounds of (\ref{eq:bound}), we propose to approximate the value of $O(s_{i},s'_{i+1})$ by:

\begin{equation}
\label{eq:approx}
O(s_{i},s'_{i+1}) \approx  \frac{d(s_i;\Omega)+d(s_{i+1};\Omega)}{2} ||s_{i+1}-s_{i}||
\end{equation}

As mentioned, the bounds for the distance function apply to obstacles that can be represented by a convex set. In general, a scenario will, however, consist of non-convex obstacles. But it is known that a non-convex set can be represented in general as the union of convex sets \cite{ORourke1983, Wei_2022}, and each of them will define a Voronoi region in which the distance function $d(l(t);\Omega)$ is convex. Therefore, the approximation in (\ref{eq:approx}) will be valid except for the segments that cross the boundaries between these regions, the cut locus.  


\revJAC{Once the derivation of (\ref{eq:approx}) is established, we present an analysis of how the line of sight, or length of the segments, affects the difference between the bounds in (\ref{eq:bound}). The difference between the bounds of (\ref{eq:bound}) is $\Delta O=||s_{i+1}-s_{i}||^2/2$ or $\Delta O$=$L^2/2$, representing the maximum deviation between the true integral and the upper bound. Considering a relative error with respect to the upper bound, suitable when obstacles are represented as unions of convex sets, we have:}

\revJAC{
\begin{equation}
\text{error}_{\text{relative}}
\approx \frac{\Delta O}{O(s_i,s_{i+1})}
= \frac{L^{2}/2}{\left(\frac{d(s_i)+d(s_{i+1})}{2}\right)L}
= \frac{L}{d(s_i)+d(s_{i+1})}.
\end{equation}
}

\revJAC{Thus, the relative uncertainty of the approximation increases proportionally with \textit{L} for fixed d(si)+d(si+1), reducing the reliability of the EDF-based term for long segments. This also implies that \textit{L} is an important parameter, as longer segments not only lead to larger approximation errors but also increase the likelihood of crossing the cut locus}. This demonstration is relevant because the value of $L$ depends on the line of sight considered, that is, the distance between $s_i$ and $s_{i+1}$ nodes will always be less than or equal to the maximum line of sight considered. So, the shorter the line of sight or length of the segments, the more similar the bounds are. This will be evaluated in Section \ref{sec:exp}.

\revJAC{Additionally, the connection with the modification introduced in Section \ref{planners} regarding limiting the line of sight can also be understood from the behavior of the EDF-based cost term in (\ref{eq:new_cost}). Since in (\ref{eq:new_cost}) the EDF-based cost appears in the denominator, the safety-related penalty term is approximately:}

\revJAC{\begin{equation}
\frac{c_w}{O(s_i,s_{i+1})}
\approx 
\frac{c_w}{\left(\frac{d(s_i)+d(s_{i+1})}{2}\right)L}
=
\frac{2c_w}{(d(s_i)+d(s_{i+1}))\,L}.
\end{equation}}

\revJAC{Therefore, for fixed values of \(d(s_i)\) and \(d(s_{i+1})\), the contribution of the EDF-based cost term scales as \(1/L\). This implies that as \(L\) increases, the penalty associated with proximity to obstacles decreases inversely with segment length, causing the Euclidean distance to dominate the total cost \(c(s_i,s_{i+1})\) in (2). Consequently, the influence of the EDF term diminishes for long segments, reinforcing the need to limit the line of sight length so that the safety contribution remains meaningful within the cost function.}

\subsection{Triangle inequality}
\label{sec:annex_ine}

\begin{figure}[!t]
	\centering
	\includegraphics[width=.79\textwidth]{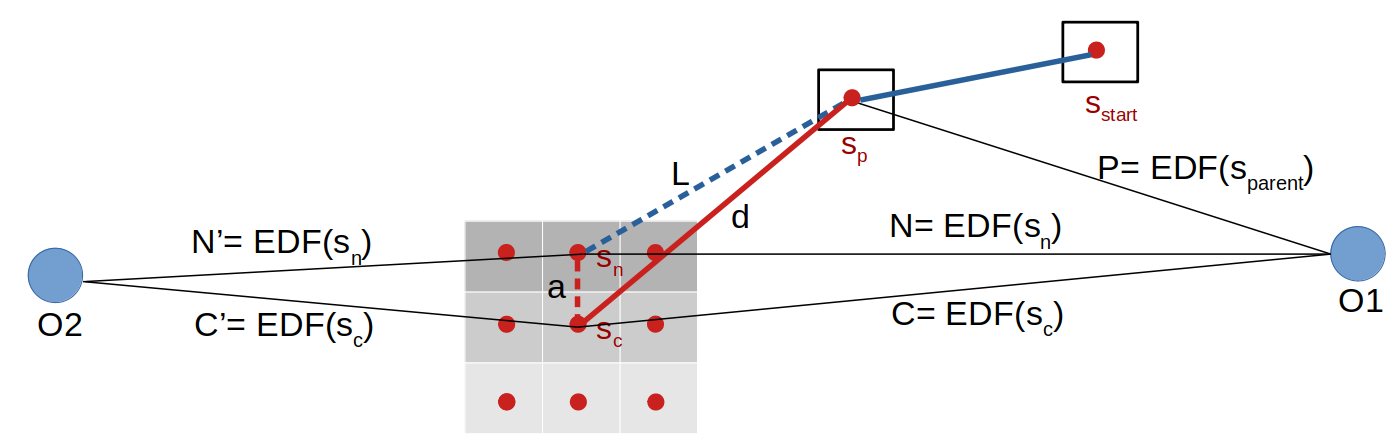}
	\caption{Computation of $g(s_n)$, $s_n$ is a neighbour node, through the parent node, $s_p$, and the current node, $s_c$, considering the distances between nodes and the EDF cost of each node. $L$ is the line of sight or distance between $s_p$ and $s_n$, $d$ is the distance between $s_c$ and $s_n$, and $a$ is the resolution of the grid or distance between $s_c$ and $s_n$. Note that distance can be $a$ or $\sqrt(2) \cdot a$}
	\label{fig:triangle}
\end{figure}

The Lazy Theta* algorithm (and many other heuristic planners) builds on top the triangle inequality \cite{THETA} that influences on the selection of the parent node. The algorithm selects a parent node of $s_n$ between $s_c$ and $s_p$ (see Figure \ref{fig:triangle}).
This selection depends only on whether there is line of sight or not, not the cost, $g$, through them. Therefore, as long as the inequality is satisfied, the path from $s_p$ to $s_n$ in Fig. \ref{fig:triangle} can be chosen over path from $s_p$ to $s_n$ through $s_c$ if $s_n$ has line of sight to the parent $s_p$. If the triangle inequality is not satisfied, both path-costs have to be explicitly computed and compared to select the parent node, which increases the computation time.

\textbf{Lemma/Proposition} Let (\ref{eq:new_cost}) be the cost function of the FS-Planner. Taking Fig. \ref{fig:triangle}, the triangle inequality is held if the cost from $s_p$ to $s_n$, $g_1(s_n)$, and the cost $s_p$ to $s_c$ to $s_n$, $g_2(s_n)$, fulfill that:
\begin{equation}
    g_1(s_n) < g_2(s_n) 
\end{equation}

\textbf{Proof} We can compute $g_1(s_n)$ and $g_2(s_n)$ from (\ref{eq:cost_function}), (\ref{eq:new_cost}) and (\ref{eq:approx}) as follows:

\begin{equation}
\label{eq:cost_parent}
    g_1(s_n)=g(s_{p})+L+\frac{2}{R \cdot L}
\end{equation}
\begin{equation}
\label{eq:cost_current}
    g_2(s_n)=g(s_{c})+a+\frac{2}{T \cdot a}=g(s_{p})+d+\frac{2}{S \cdot d}+a+\frac{2}{T \cdot a}
\end{equation}

\noindent where $R=P+N'$, $S=P+C'$, $T=C'+N'$ and $P, N', C'$ are the value of the EDF for $s_p, s_n, s_c$, respectively (see Fig. \ref{fig:triangle}).

Figure \ref{fig:triangle} shows cases where the obstacle $O1$ is the closest to $s_{p}$, $s_c$ and $s_n$ or $O2$ is closer than $O1$ to $s_c$ and $s_n$. We consider the second one because it encompasses all possible cases. The EDF is Lipstchiz continuous, with the Lipstchiz constant equal to $1$, and we can ensure that the difference between the EDF cost of two nodes is always smaller than the distance between the nodes \cite{Jones:TVCG2006}. Therefore, $|P-N'|<L$, $|P-C'|<d$ and $|N'-C'|<a$.

Note that $L<d+a$ due to triangle inequality of the Euclidean distance, so we can discard the terms $g(s_p)$, $L$, $d$ and $a$ in (\ref{eq:cost_parent}) and (\ref{eq:cost_current}), and verify that the last term in (\ref{eq:cost_parent}) is smaller than the third and fifth terms in (\ref{eq:cost_current}) and thus the triangle inequality will be held:
\begin{equation}
\label{eq:inequality1}
    \frac{2}{R \cdot L} < \frac{2 \cdot S \cdot d + 2 \cdot T \cdot a}{S \cdot T \cdot d \cdot a}
\end{equation}

\noindent This inequality will always be satisfied if the minimum value of the right term is higher than the maximum value of the left term of (\ref{eq:inequality1}). Taking into account the EDF properties presented above, we obtain $S-T=|P-N'| \rightarrow S-T<L $. Therefore, $S$ is upper bounded by $L+T$. We can also consider that $S\cdot T < L \cdot T + T^2$ as $S < L+T$. Next we have to check if this value ensures the minimum value of the term on the right of (\ref{eq:inequality1}). The first derivative of the fraction with respect to $S$ is:
\begin{equation}
\label{eq:first_der}
    f'(S) = - \frac{2}{S^2 \cdot d}
\end{equation}

As the first derivative is negative, the function is decreasing, so the minimum value of (\ref{eq:first_der}) is reached with the maximum value of $S$. Regarding the term on the left of (\ref{eq:inequality1}), we know that $|R-T|<d \rightarrow R=T+d$. In order to consider the smallest value of R, we consider $R=T+a$ as $d \geq a$. Therefore, the inequality follows as:
\begin{equation}
\label{eq:inequality2}
    \frac{2}{T \cdot L + L \cdot a} < \frac{2 \cdot L \cdot d + 2 \cdot T \cdot d + 2 \cdot T \cdot a }{L \cdot T \cdot d \cdot a + T^2 \cdot d \cdot a}
\end{equation}

In order to calculate the smallest term we have to multiply each numerator by the other denominator:
\begin{multline}
\label{eq:inequality3}
    2 \cdot L \cdot T \cdot d \cdot a + 2 \cdot T^2 \cdot d \cdot a < 2 \cdot L^2 \cdot T \cdot d + 2 \cdot  T^2 \cdot L \cdot d \\ + 2 \cdot L \cdot T^2 \cdot a + 2 \cdot L^2 \cdot d \cdot a + 2 \cdot L \cdot T \cdot d \cdot a + 2 \cdot L \cdot T \cdot d \cdot a
\end{multline}

Note that the term $2 \cdot L \cdot T \cdot d \cdot a$ appears in both sides, so it can be simplified. As $L \geq a$, the terms $2 \cdot T^2 \cdot d \cdot a$ and $2 \cdot T^2 \cdot L \cdot d$ can also be simplified. Thus, it is demonstrated that the last term in (\ref{eq:cost_parent}) is always smaller than the third and fifth terms in (\ref{eq:cost_current}). Therefore, the triangle inequality is held.

\section{Selection of visibility neighbours}
\label{selection}

This section explains how the selection of visibility neighbours from the obstacle-gradient information of the EDF representation and its properties \cite{Jones:TVCG2006} is performed with the aim of generating fast and safe paths. Unlike Lazy Theta*, our approach does not evaluate all possible neighbours around the current node. Instead, we implement a neighbour selection scheme that is intrinsically safe in terms of distance to obstacle, while reducing the number of explored neighbours. For this, we select the neighbours that imply moving away from the obstacle thanks to the obstacle-gradient information and thus safer paths are computed. 

\begin{figure*}[t]
    \centering
    \begin{subfigure}[b]{0.63\textwidth}
        \centering
        \includegraphics[width=\linewidth]{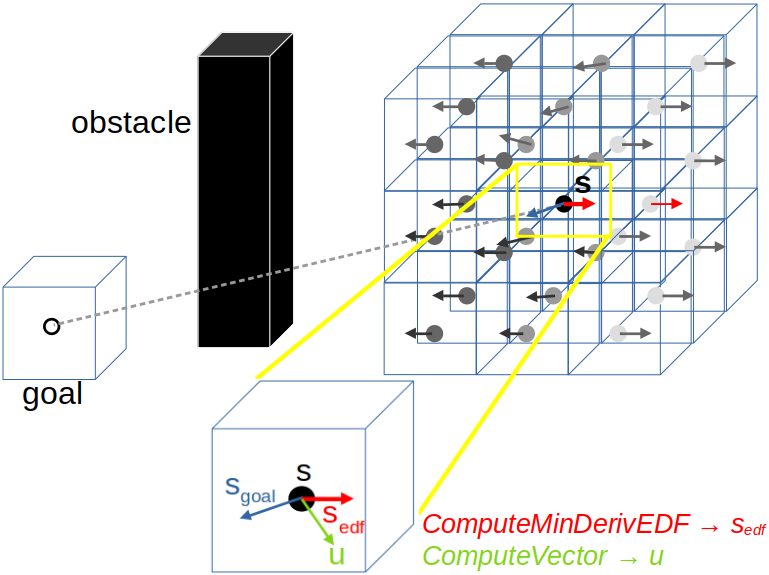}
        \caption{}
        \label{fig:choose_neigh_a}
    \end{subfigure}
    \hfill
    \begin{subfigure}[b]{0.35\textwidth}
        \centering
        \includegraphics[width=\linewidth]{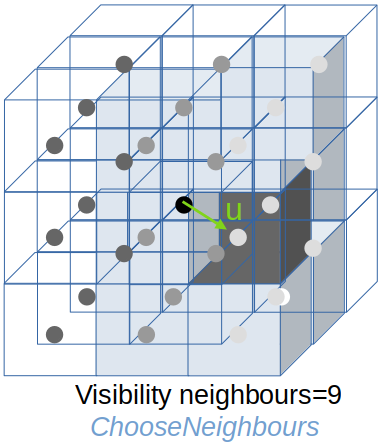}
        \caption{}
        \label{fig:choose_neigh_b}
    \end{subfigure}
    \caption{\revJAC{Visibility-neighbour selection process. (a) The neighbour with the minimum value of $EDF'(s')$ is identified ($s_{edf}$) by the \textit{ComputeMinDerivEDF} function, indicated by the two red arrows, together with the direction to the goal $s_goal$ (blue arrow). These two vectors are combined by the \textit{ComputeVector} function to obtain the unit direction vector $\hat{u}$ (black arrow). (b) Once the best candidate neighbour (dark gray cell) is selected, the \textit{ChooseNeighbours} function generates the reduced set of visibility neighbours; in this example, nine neighbours are selected (blue cells).}}
    \label{fig:choose_neigh}
\end{figure*}

First, the directional derivative of the $EDF$, $EDF'$, in the direction of each neighbour of the twenty six, $s'$, from the current node, $s$, in 3D is computed as follows (see red arrows in Figure \ref{fig:choose_neigh}(a)): 
\begin{equation}
\label{eq:grad}
    EDF'(s)=\frac{EDF(s)-EDF(s')}{distance(s,s')}
\end{equation}

\noindent where $EDF(s)$ and $EDF(s')$ are the distances to the closest obstacles in each node $s$ and $s'$, 
and $distance$ is the distance between $s$ and $s'$. Notice how positive values indicate that we are getting closer to an obstacle in such direction, while negative ones imply moving away from the obstacle.

Then, the minimum value of $EDF'(s')$ of the twenty six neighbours, $s_{edf}$, is chosen with the $ComputeMinDerivEDF$ function in Algorithm \ref{lazy_proposed} (see red arrow in Figure \ref{fig:choose_neigh_a}). On the other hand, the direction of $s$ to the goal is also computed 
(blue arrow in Figure \ref{fig:choose_neigh_a}).
\revJAC{Once $s_{edf}$ and $s_{goal}$ are computed, the $ComputeVector$ function in line 16 of Algorithm \ref{lazy_proposed} calculates the sum of their corresponding direction vectors and normalizes the result to obtain a unit vector $\hat{u}$ (green arrow in the enlarged part of Figure \ref{fig:choose_neigh_a}). This vector defines the direction toward the best candidate node among the 26 visibility neighbours to be expanded}
(green arrow is the direction and \revJAC{dark gray cell} the best candidate to expand in Figure \ref{fig:choose_neigh_b}). 
\revJAC{Following the computation of $\hat{u}$, $ChooseNeighbours$ function in line 17 in Algorithm \ref{lazy_proposed} selects the neighbours around the best candidate (blue cells in Figure \ref{fig:choose_neigh_b}). }
For example, \revJAC{if the \method is configured to explore 9 neighbours out of the 26 surrounding node $s$, the vector $\hat{u}$ identifies the best candidate to expand, and the remaining 8 neighbours correspond to the cells surrounding this candidate (dar gray cell and blue cells in Figure \ref{fig:choose_neigh_a}.} Figure \ref{fig:choose_neigh} \revJAC{illustrates the selection process when 9 neighbours are chosen for expansion}. Finally, the cost, $g$, and parent of each node is computed for the selected neighbours in line 18-23 in Algorithm \ref{lazy_proposed}. 

This new neighbour selection policy significantly improves search efficiency, as the number of expanded nodes is dramatically decreased. Obviously, the proposed algorithm computes a sub-optimal path because all nodes are not expanded. Nevertheless, experiments will show that the solutions are very close to the optimal. 
In order to ensure the completeness of the algorithm, if the open list becomes empty \cite{Hart1968}, the {\method} will be run again considering all the neighbours as in \cite{iros2022_edf} to ensure the solution. 


\subsection{Case study}

\begin{figure*}[t]
    \centering
    \begin{subfigure}[b]{0.43\textwidth}
        \centering
        \includegraphics[width=\linewidth]{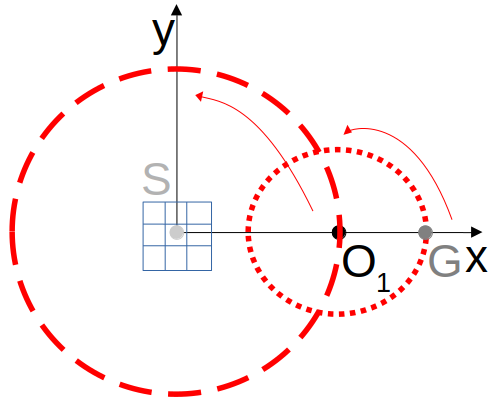}
        \caption{}
        \label{fig:quality_a}
    \end{subfigure}
    \hfill
    \begin{subfigure}[b]{0.48\textwidth}
        \centering
        \includegraphics[width=\linewidth]{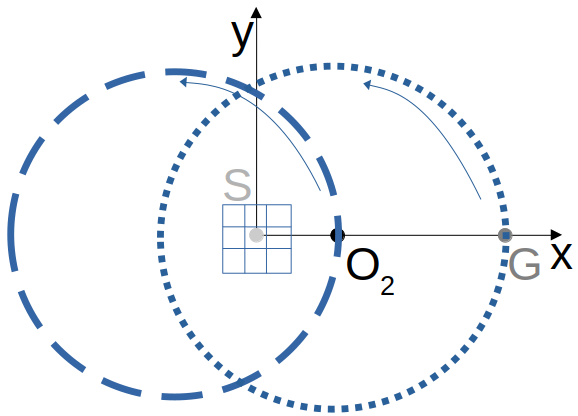}
        \caption{}
        \label{fig:quality_b}
    \end{subfigure}
    \caption{Selection of neighbours in 2D. The current node\revJAC{, $S$,} is in the coordinates' center, $O_1$ and $O_2$ are the obstacles, and $G$ the goal. The grid in the centre shows the eight neighbours of the current node. \revJAC{As an analogy with the Sun–Earth–Moon system, in each case $S$ can be viewed as the Sun, $O_1$ and $O_2$ as the Earth, and $G$ as the Moon. Thus, points $O_1$ and $O_2$ rotate around $S$ like the Earth around the Sun, while G rotate around O1 and O2 in each case, respectively, like the Moon around the Earth}. First case (a): $G$ rotates with a small radius around $O_1$ (dotted line in red). Second case (b): $G$ rotates around $O_2$ (dotted line in blue) with a large radius. The difference between both setups is that in the second one the current node might fall between $G$ and $O_2$.}
    \label{fig:quality}
\end{figure*}

This section presents a study to evaluate the quality of the exploration in the {\method}, as it does not guarantee an optimal path because it only expands selected nodes. We will define the quality of selection as the relation between the selected nodes among all the neighbours and the order of these nodes ranked by their estimated cost $g+h$. For simplicity, we analyze the 2D case where eight neighbours are explored. Figure \ref{fig:quality} shows two cases by varying the obstacle and goal to consider all possibilities depending on the current node, obstacle, and goal. In the second case, we can analyze cases where the current node is between the obstacle and the goal. 

\begin{figure}[!t]
	\centering
    \includegraphics[width=0.79\textwidth]{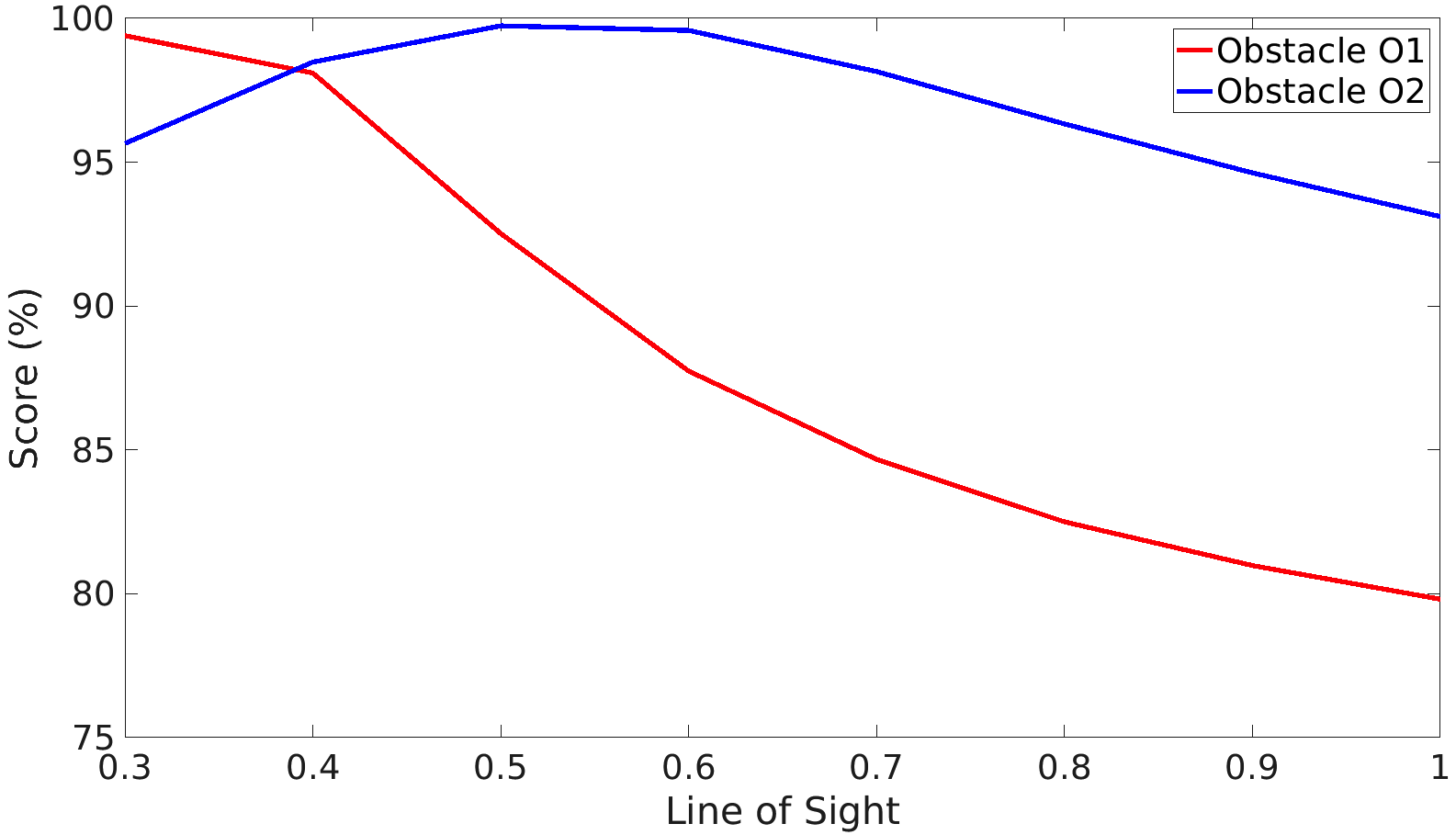}
	\caption{Percent of the quality in the tests performed when three nodes are chosen.}
	\label{fig:analysis}
\end{figure}

Let us consider the selection of three and five visibility neighbours out of the eight possible ones in 2D by considering the cases of Fig. \ref{fig:quality} and from the obstacle-gradient information (Section \ref{selection}). In order to measure the quality of the selection, we sort the eight neighbour nodes according to their cost from lowest to highest in each exploration, and analyze which of them are selected. The analysis focuses on determining the percentage of solutions in which at least one of the three lowest-cost nodes is selected among the neighbours selected by our approach: a score of $100\%$ indicates that at least one of the three lowest-cost nodes is in the neighbours selected by our approach in the 100\%  of the tests. Figure \ref{fig:analysis} shows the results obtained when three neightbours are selected for the two cases presented in Fig. \ref{fig:quality}. We can see how the score varies on the line of sight and how, in general, is high, while the minimum score is $79.81\%$. This means that only expanding the three nodes selected by our approach, we are able to get sub-optimal but similar results to expanding all nodes in most cases.

We repeated the same experiment, but selecting five neighbours instead of three. In this case, the score is $100\%$ regardless the line of sight (we do not show in Fig. \ref{fig:analysis} for simplicity) in both cases. This means that no matter the line of sight, our approach always selects at least one of the lowest-cost nodes if five neighbours are selected. Thus, five neighbours improve the results w.r.t. three, but at the cost of expanding $40\%$ more nodes.

\section{Experimental results} \label{sec:exp}

This section presents experimental results in different 3D scenarios in order to analyze the robustness and performance of the {\method}. The method is benchmarked by considering different number of the visibility neighbours selected among all the nodes. The {\method} is tested with different numbers of visibility neighbours in 3D: $9, 10, 11, 13, 15$ or $17$ nodes. The proportion of nodes selected relative to all neighbours is similar to the study in 2D ($3$ or $5$ out of $8$), that is, from $9$ to $17$ out of $26$. In all the cases, the neighbours are selected around the best candidate, $\hat{u}$, (see Section \ref{selection}) except when $10$ nodes are selected. In this particular case, we consider $9$ nodes around the best candidate, $\hat{u}$, and the node opposite to it. A third method is implemented depending on the angle between the minimum gradient unit vector, the unit vectors of $s_{edf}$, and the goal unit vector, the unit vector of $s_{goal}$. If the angle is smaller than $90^o$ we consider a number of neighbours and if the angle is greater than $90^o$ a larger number of neighbours is considered. For example, $9-11$ considers $9$ neighbours when the angle is smaller than $90^o$.

We will benchmark the {\method} against the CALT*+M1 planner presented in \cite{iros2022_edf}, and the A* algorithms. All planners compute the path on the same EDF previously generated, so the computation times are exclusively referred to the generation of the path. Both FS-Planner and A* make use of the same EDF-based cost function (\ref{eq:new_cost}). As A*+M1 in \cite{iros2022_edf} mentioned in Section \ref{sec:intro}, CALT*+M1 considers an EDF as environment representation and a cost function that integrates costs related to the distance to obstacles. The results in \cite{iros2022_edf} showed that the CALT*+M1 provided similar computation times to A*, while safety \revJACS{ and heading changes between consecutive segments of the path} are clearly improved. Moreover, A* algorithm is used as baseline in order to compare them considering the three algorithms the same cost function (\ref{eq:new_cost}).

\begin{figure}[t!]
  \hspace{20mm}
  \begin{subfigure}[b]{0.23\linewidth}
    \centering
    \includegraphics[width=\linewidth]{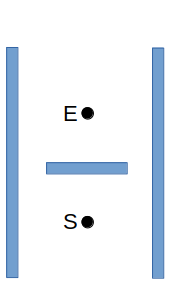}
    \caption{}
    \label{fig:sce1}
  \end{subfigure}\hspace{20mm}
  \begin{subfigure}[b]{0.23\linewidth}
    \centering
    \includegraphics[width=\linewidth]{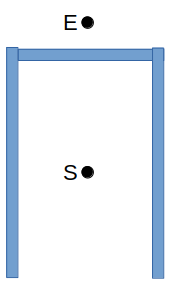}
    \caption{}
    \label{fig:sce2}
  \end{subfigure}

  \vspace{1em} 

  \hspace{20mm}
  \begin{subfigure}[b]{0.23\linewidth}
    \centering
    \includegraphics[width=\linewidth]{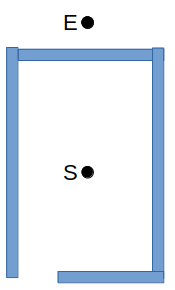}
    \caption{}
    \label{fig:sce3}
  \end{subfigure}\hspace{10mm}
  \begin{subfigure}[b]{0.38\linewidth}
    \centering
    \includegraphics[width=\linewidth]{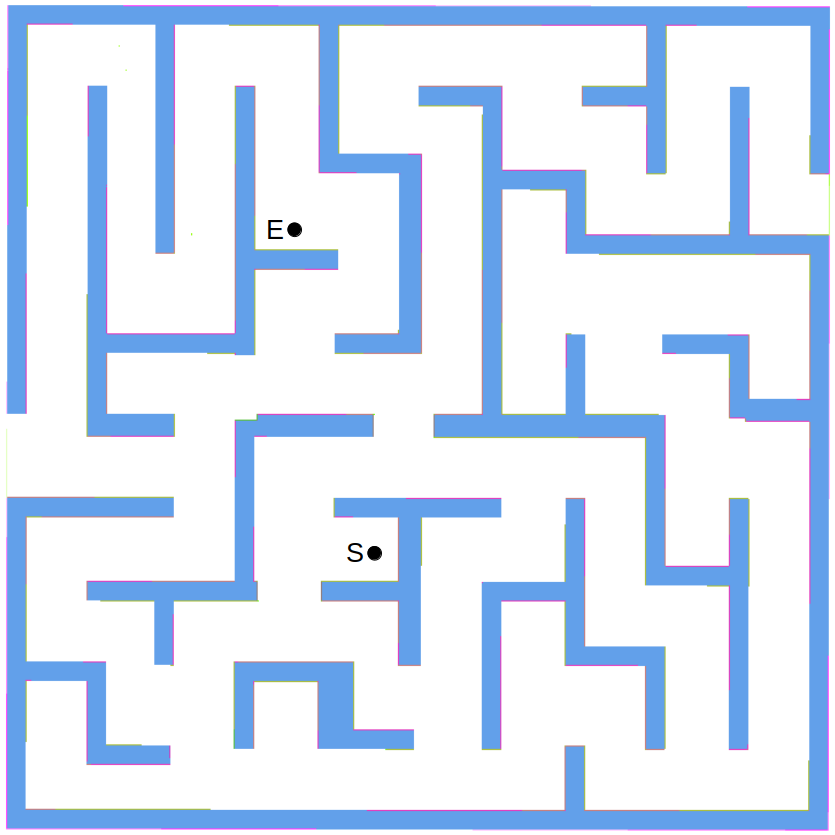}
    \caption{}
    \label{fig:sce4}
  \end{subfigure}
  
  \caption{Scenarios considered in the tests performed: \revJAC{(a) S1, (b) S2, (c) S3 and (d) S4.} \revJAC{S is the start point and E the end point of the paths considered in each scenario.}}
  \label{fig:sce}
\end{figure}

\begin{figure*}[t]
    \centering
    \begin{subfigure}[b]{0.6\textwidth}
        \centering
        \includegraphics[width=\linewidth]{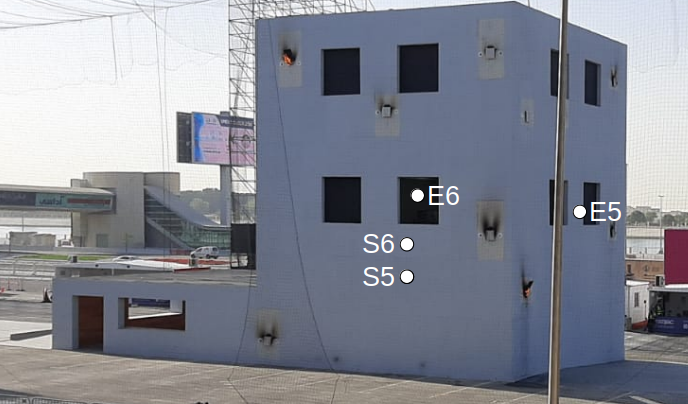}
        \caption{}
        \label{fig:mbzirc}
    \end{subfigure}
    \hfill
    \begin{subfigure}[b]{0.45\textwidth}
        \centering
        \includegraphics[width=\linewidth]{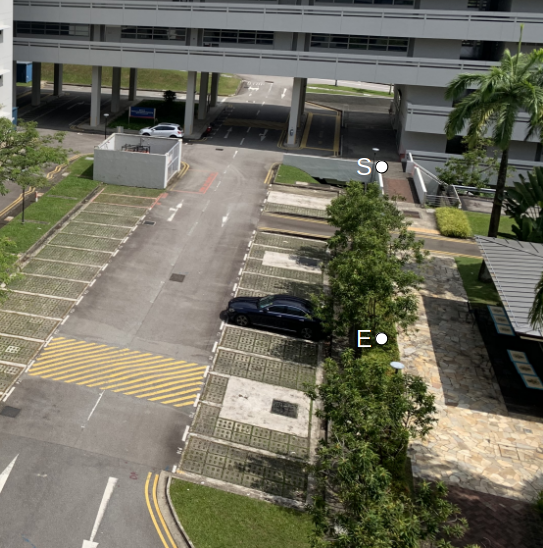}
        \caption{}
        \label{fig:viral}
    \end{subfigure}
    \hfill
    \begin{subfigure}[b]{0.48\textwidth}
        \centering
        \includegraphics[width=\linewidth]{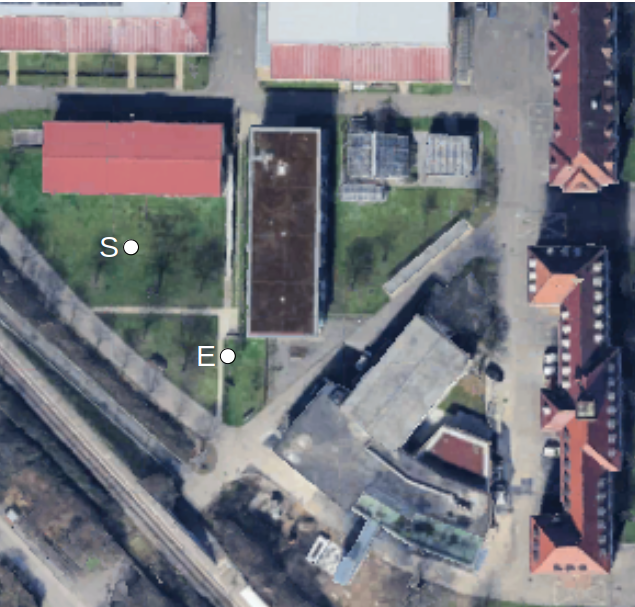}
        \caption{}
        \label{fig:freiburg}
    \end{subfigure}
    \caption{\revJACS{Real-world outdoor environments considered in the tests performed: (a) S5 and S6, (b) S7 and (c) S8. S is the start point and E the end point of the paths considered in each scenario.}}
    \label{fig:real_sce}
\end{figure*}

The \revJACS{simulated} scenarios \revJACS{generated to test the planners} are presented in Fig. \ref{fig:sce}: S1, like letter H, (see Fig. \ref{fig:sce1}); S2, like letter U inverted, (see Fig. \ref{fig:sce2}); S3 (see Fig. \ref{fig:sce3}); S4 is the maze shown in Fig. \ref{fig:sce4}). \revJACS{The tests have also been carried out in more realistic environments and are presented in Fig. \ref{fig:real_sce}:} S5 \revJACS{and S6 consider} the same building \revJACS{used} in \cite{iros2022_edf} ( \revJACS{see Fig. \ref{fig:mbzirc},} \ref{fig:explored_nodes} \revJACS{and \ref{fig:paths_building}})\revJACS{. Figure \ref{fig:real_sce} also shows the real-world outdoor environments used to test the planners; S7 considers point clouds derived from real-world data collected in an outdoor environment on the Nanyang Technological University (NTU) campus, referred to as the EEE environment in \cite{nguyen2022ntu} (see Fig. \ref{fig:viral}); S8 considers the point clouds derived from real data in the Computer Science campus at University of Freiburg\footnote{http://ais.informatik.uni-freiburg.de/projects/datasets/octomap/} (see Fig. \ref{fig:freiburg})}. The objective is to test the {\method} in challenging scenarios, with \revJACS{buildings,} narrow openings\revJACS{, trees and} windows considering convex and non-convex obstacles. \revJACS{Note that we consider two paths in the scenario of \cite{iros2022_edf} (see Fig. \ref{fig:mbzirc}): the first one goes from one building facade to another turning in a corner defined by S5 and E5 points (see Fig. \ref{fig:mbzirc}, and Fig. \ref{fig:paths_building}); and the second one passes through the window from the outside to the inside defined by S6 and E6 poinsts (see Fig. \ref{fig:mbzirc}). Regarding S7 and S8, it is important to note that the start (S) and end (E) points are located close to ground level and below the tree canopies. This aspect can be more clearly appreciated in the corresponding path visualizations shown in Fig. \ref{fig:paths_viral} and \ref{fig:paths_campus}}



The computer used to perform the tests \revJAC{is}: Intel Core i7-10700 CPU 2.90GHz, 32 GB RAM, with Ubuntu OS 20.04 LTS. All the planning algorithms have been developed in C++ language and has been integrated into ROS (Robot Operating System) Noetic distribution.

\subsection{Dependency on the number of visibility neighbours}
\label{dependency}

This section shows the performance of the {\method} depending on the number of visibility neighbours and how it always outperforms CALT*+M1. We focus on the computation time and the explored nodes. \revJAC{We evaluate different numbers of selected neighbours to characterize the behavior of the {\method}. Since the {\method} modifies the original Lazy Theta* through the introduction of a gradient-based neighbour-selection mechanism, it is important to show how this mechanism behaves as the number of visibility neighbours changes.}

\begin{figure}[!t]
	\centering
    \includegraphics[width=.99\textwidth]{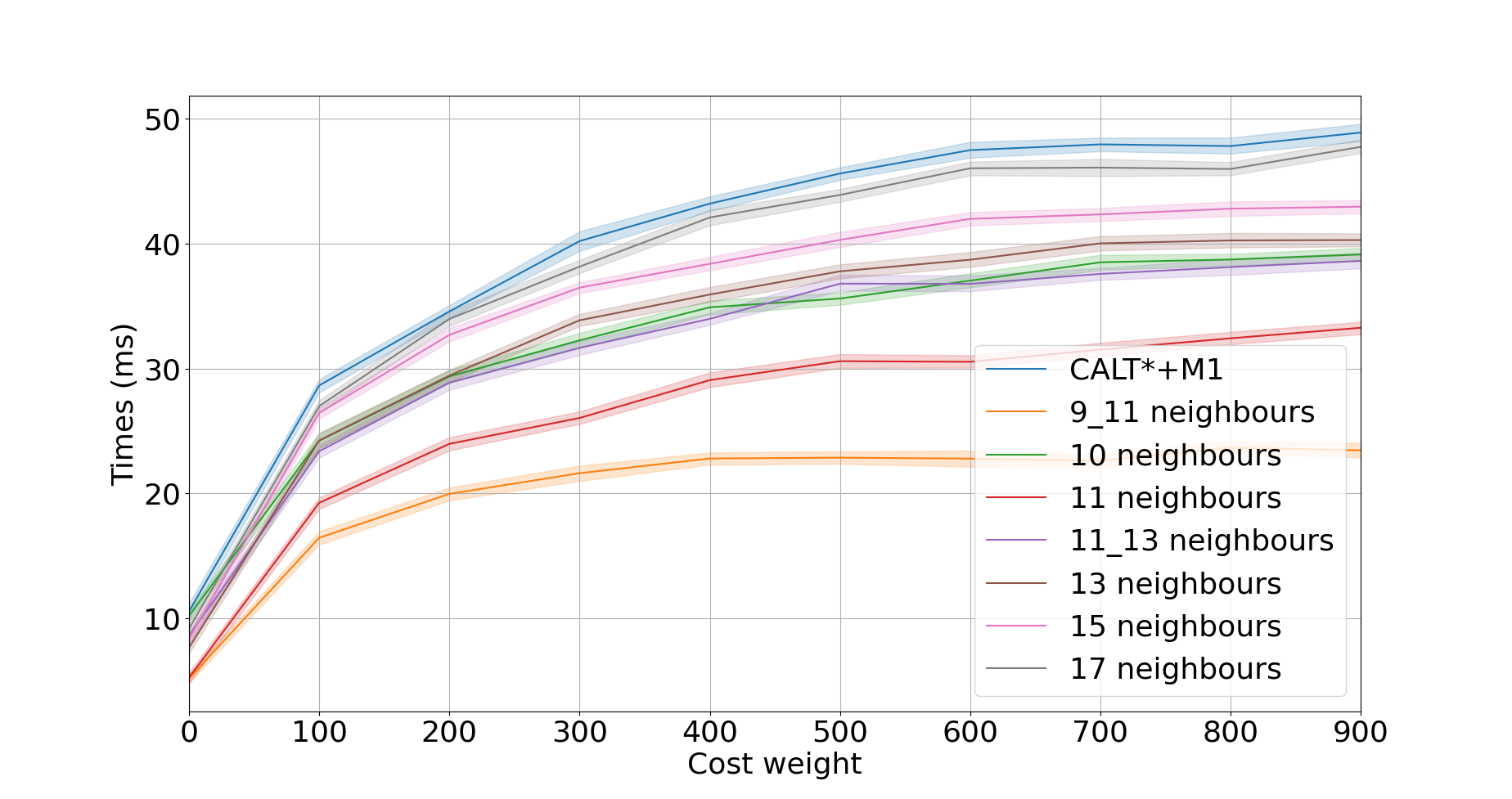}
 \caption{Computation time in S1 considering CALT*+M1 and {\method} with a different number of selected neighbours:  9, 10, 11, 13, 15 or 17 node. The path goes from 1 to 3 points and $LoS=1$.}
	\label{fig:h_times}
\end{figure}

Figure \ref{fig:h_times} shows the mean computation times and the $95\%$ confidence interval in scenario S1. We consider different values of $c_w$ for the CALT*+M1 
and the {\method}. 
We can see how the time increases with $cw$ but it remains almost constant from $cw=500$. Notice how the computation time of {\method} is always shorter than CALT*+M1 no matter the number of neighbours selected.

\begin{figure*}[t]
    \centering
    \begin{subfigure}[b]{0.48\textwidth}
        \centering
        \includegraphics[width=\linewidth]{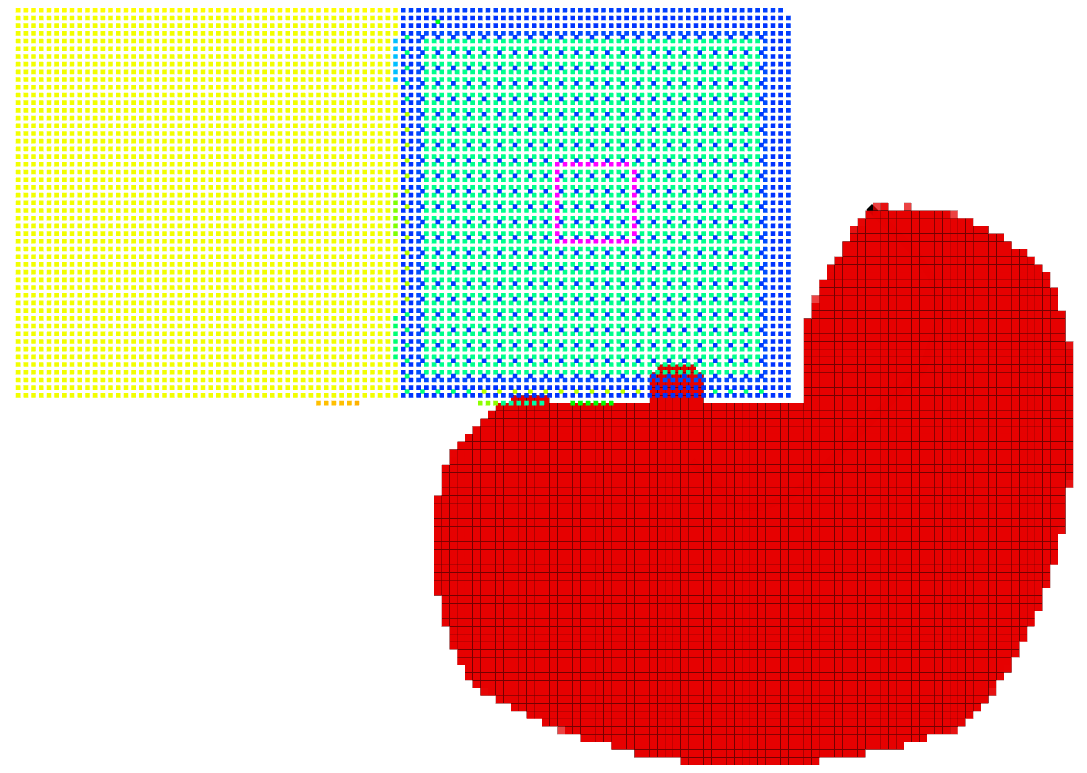}
        \caption{}
        \label{fig:quality_a}
    \end{subfigure}
    \hfill
    \begin{subfigure}[b]{0.48\textwidth}
        \centering
        \includegraphics[width=\linewidth]{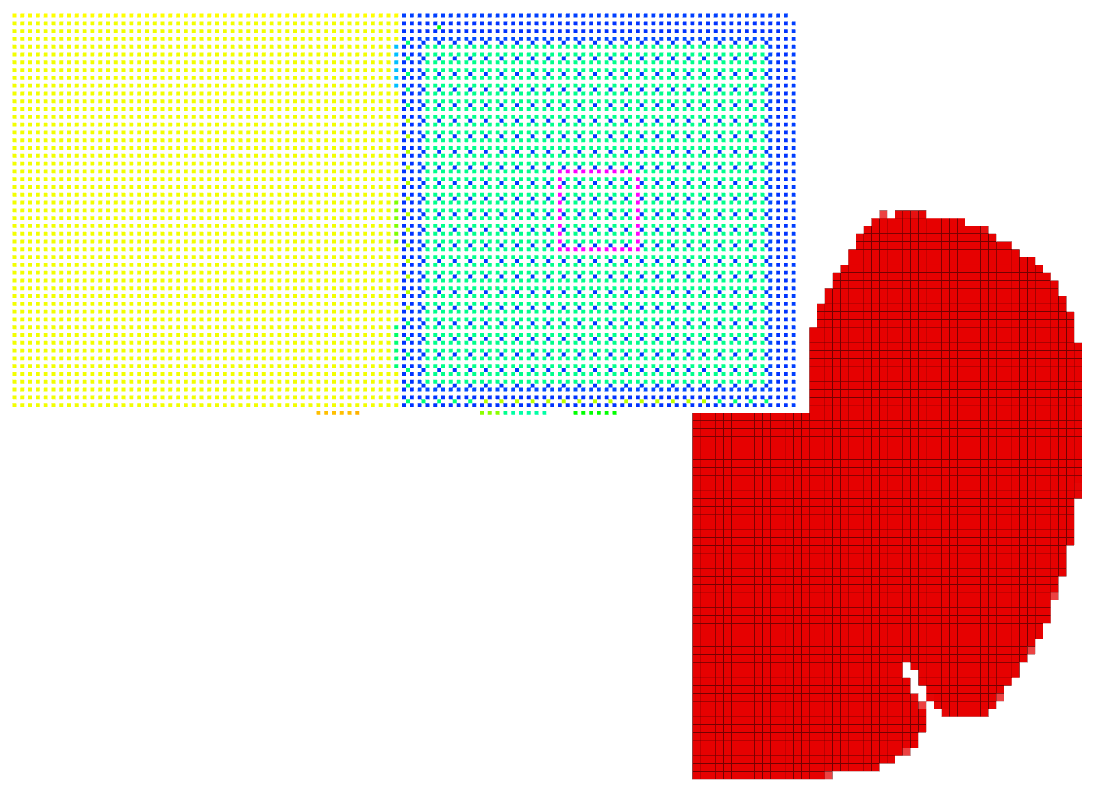}
        \caption{}
        \label{fig:quality_b}
    \end{subfigure}
    \caption{Explored nodes in red to compute a path that goes from one building facade to another turning in a corner with: (a) CALT*+M1 and (b) {\method} considering 11 visibility neighbours.}
    \label{fig:explored_nodes}
\end{figure*}

\revJAC{Figure \ref{fig:explored_nodes} illustrates how the gradient-based visibility neighbour selection significantly reduces the number of explored nodes in the {\method} compared to CALT*+M1. The path considered corresponds to the first path in scenario S5, where the trajectory moves from one building façade to another while turning around a corner (see Figure \ref{fig:paths_maze}-(right)). Note that CALT*+M1 explores a portion of the building’s interior, whereas the {\method}, by incorporating gradient information, avoids this exploration and focuses the search on the relevant region of the environment.}


\subsection{Random tests}
\label{random}

This section presents an evaluation with random tests to compare the CALT*+M1 and the {\method} using the A* algorithm as baseline. All planners make use of a previously computed EDF. Both {\method} and A* use the same cost function (\ref{eq:new_cost}). Note that we are interested in the computation times of each planner on EDF and not in the EDF generation. 
\revJAC{The metrics used to evaluate efficiency, safety, and smoothness follow the definitions in \cite{iros2022_edf} (see Table \ref{table_results}): path length, computation time and number of explored nodes (related to efficiency), clearance to obstacles (related to safety), and} \revJACS{a metric of geometric smoothness based on heading changes between consecutive path segments.} \revJAC{For each scenario, we perform twenty planning tests with randomly distributed start and goal positions. }Twenty simulations are performed in each problem to calculate the mean computation time. It is considered $c_w=500$ and $LoS=1$ in all the cases. Note that the difference between the bounds of (\ref{eq:bound}) is small ($0.5$), so using the upper bound in (\ref{eq:approx} provides a good approximation for both convex and non-convex shapes. As a result, we do not need to infer the type of obstacle. The number of visibility neighbours are: 9-11, 10, 11 and 11-13. 
The solution is always found in the tests shown in this section, but 
the implementation of the {\method} also considers the case that no solution is found. In that case the {\method} will be run again considering $17$ neighbours to ensure the solution.


Table \ref{table_results} summarises the \revJAC{main results for each scenario using} A* algorithm\revJAC{, considering (\ref{eq:new_cost}),} as baseline  \revJAC{for comparison}. \revJAC{For each metric, the} ratio is calculated as $\frac{algorithm}{A*}$, \revJAC{allowing a direct assessment of the relative performance with respect to the baseline. Table \ref{table_results} reports the mean over twenty runs together with the standard error of the mean, which quantifies the uncertainty of the estimated average. Lower values of T, L, N and MA indicate better performance, whereas higher values of MD denote safer paths}. The best values are highlighted in bold.

All the results show that the \revJAC{computation times (T) and number of explored nodes (N)} of the {\method} considering different number of visibility neighbours are lower than the ones of CALT*+M1, as in Fig. \ref{fig:h_times}. \revJAC{Particularly, the {\method} configurations with 9–11 visibility neighbours obtain the lowest computation times and the smallest number of explored nodes in all} \revJACS{eight} \revJAC{scenarios.} Even the {\method} shows times lower than A* algorithm over EDF using (\ref{eq:new_cost}) in scenarios S5 and S6 (see Table \ref{table_results}). 
The length is usually very close and similar to the one of A* with CALT*+M1 and {\method}. It is greater than the length of A* only with the {\method} \revJACS{in some cases} in S3 \revJACS{and S8}. \revJACS{S3} is really challenging and the computed path is longer inside of the U almost closed.

\begin{figure}[!t]
	\centering
    \includegraphics[width=.99\textwidth]{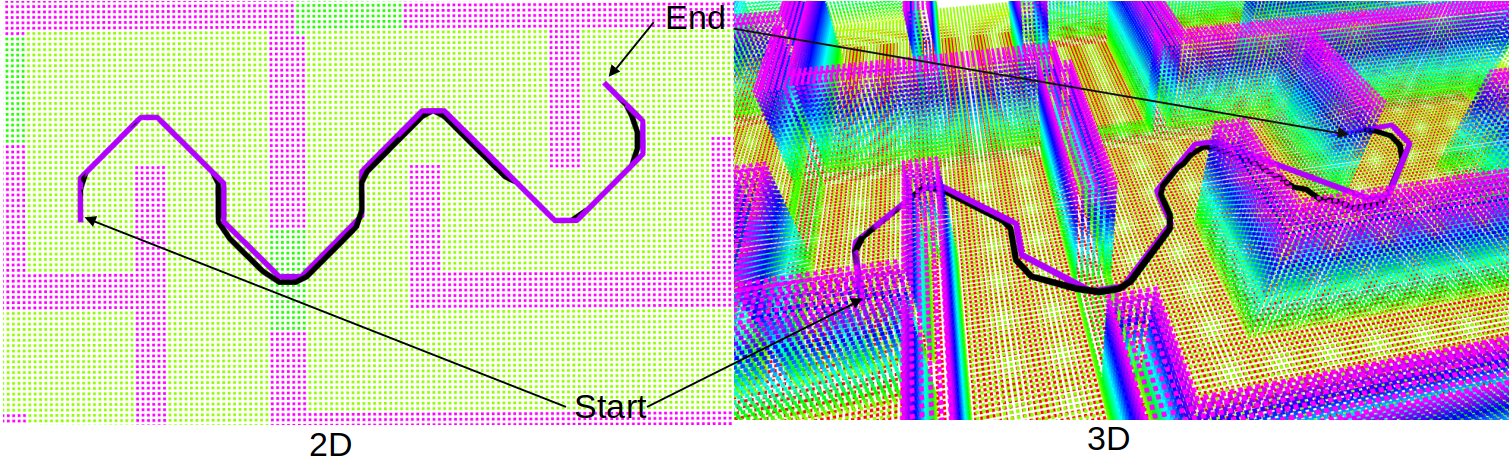}
	\caption{\revJAC{Paths considered in S4 in 2D (left) and 3D (right): A* in purple and {\method} 9-11 in black.}}
	\label{fig:paths_maze}
\end{figure}

\revJAC{Regarding safety (MD) and \revJACS{geometric} smoothness (MA), the {\method} obtains the best values in} \revJACS{six} \revJAC{out of the} \revJACS{eight} \revJAC{scenarios. In the scenario S4} \revJACS{and S8}\revJAC{, its performance is very close to that of CALT*+M1, showing that reducing the number of explored neighbours does not degrade the safety and} \revJACS{geometric} \revJAC{smoothness of the computed path.} Also it is safer than A*  in scenarios S1, S2, S3, S6 \revJACS{and S7}, and very much alike in scenarios S4, S5 \revJACS{and S8}. 
\revJAC{Figure \ref{fig:paths_maze} depicts the paths computed by A* and the {\method} in scenario S4. Both paths are safe, but the one generated by the {\method} is \revJACS{geometrically} smoother, particularly in turns and during ascending or descending segments as is shown in Table \ref{table_results}.}

\begin{figure}[!t]
	\centering
    \includegraphics[width=.99\textwidth]{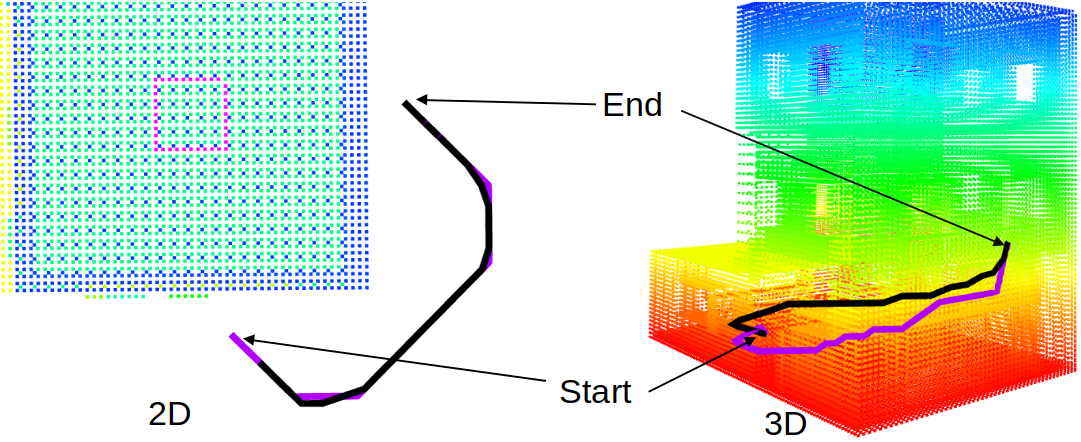}
	\caption{\revJAC{Paths considered in S5 in 2D (left) and 3D (right): A* in purple and {\method} 9-11 in black.}}
	\label{fig:paths_building}
\end{figure}

\begin{figure}[!t]
	\centering
    \includegraphics[width=.99\textwidth]{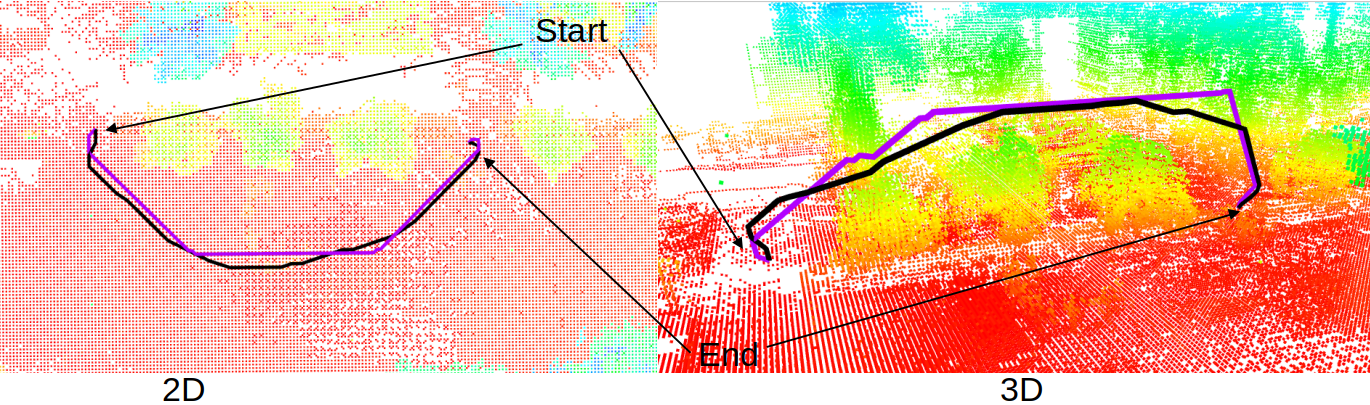}
	\caption{\revJAC{Paths considered in S7 in 2D (left) and 3D (right): A* in purple and {\method} 9-11 in black.}}
	\label{fig:paths_viral}
\end{figure}

\begin{figure}[!t]
	\centering
    \includegraphics[width=.99\textwidth]{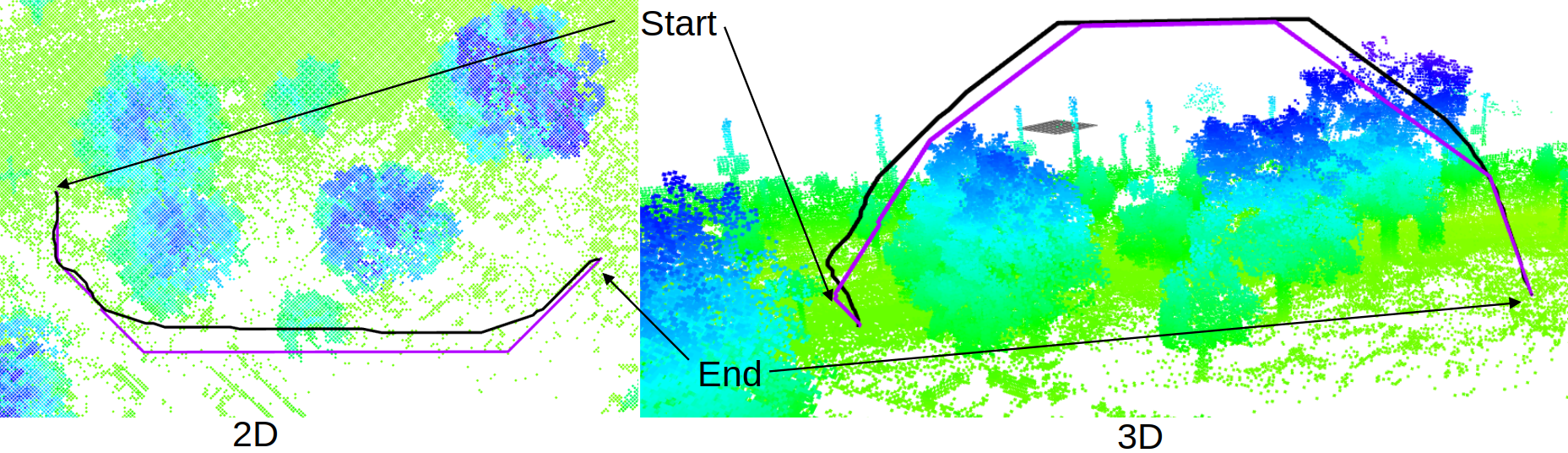}
	\caption{\revJAC{Paths considered in S8 in 2D (left) and 3D (right): A* in purple and {\method} 9-11 in black.}}
	\label{fig:paths_campus}
\end{figure}

It is important to highlight the results obtained in the real scenarios considered in \revJACS{S5, S6, S7 and S8}. 
The {\method} shows better values than CALT*+M1. Moreover, the computation times of the {\method} with 9-11 and 11 neighbours are lower than the ones with A* algorithm. The length and safety are very similar and the paths are much \revJACS{geometrically} smoother than the ones computed by A* algorithm. Figure \revJAC{\ref{fig:paths_building}} illustrates the improvement of the \revJACS{geometric} smoothness with the {\method} in S5. Note that the path computed by {\method} 9-11 is \revJACS{geometrically} smoother than the one computed by A* \revJAC{during ascending segments (see Figure \ref{fig:paths_building}-right)}. \revJACS{Regarding real-world scenarios, S7 and S8, Figures \ref{fig:paths_viral} and \ref{fig:paths_campus} also illustrate the improvement of the smoothness with the {\method}.}

The results demonstrate that the {\method}, which exploits the EDF properties and reduces the number of visibility neighbours, works better than CALT*+M1 in terms of computation times, explored nodes, safety and \revJACS{geometric} smoothness. 
Although the length is very similar in both planners, CALT*+M1 computes slightly shorter paths. 
Regarding A* algorithm over EDF using the cost function (\ref{eq:new_cost}), the \revJACS{geometric} smoothness of the path computed by {\method} is considerably improved and the times too in some scenarios.
Therefore, we can conclude the {\method} improves both CALT*+M1 and A* algorithm over EDF using cost function (\ref{eq:new_cost}). Moreover, both the {\method} and CALT*+M1 do not make use of post-processing methods to achieve \revJACS{geometrically} smoother paths.

\revJAC{On the other hand, analyzing different numbers of selected neighbours allows us to empirically assess the robustness and scalability of the {\method}, and to clarify how the number of selected neighbours influences the main performance metrics (computation time, explored nodes, safety, and \revJACS{geometric} smoothness). This evaluation confirms that the {\method} maintains high performance even with reduced neighbour sets.}

\revJACS{Finally, it also is important to note that the experimental results provide empirical evidence of the behaviour of the {\method}, but they do not constitute formal guarantees of completeness or optimality. Although the fallback mechanism, where the {\method} is executed again considering all the neighbours, was never activated in any of the tested scenarios, including complex 3D and real-world environments, this observation should not be interpreted as a proof of completeness. Instead, it indicates that the neighbour selection policy behaves robustly across all evaluated scenarios.} 

\revJACS{Moreover, the twenty random planning tests performed per scenario indicate that the main properties and behaviour of the {\method} are consistently preserved across different configurations and situations. Although this does not constitute statistically exhaustive evidence, we believe that the results are very positive in terms of the quality of the generated paths and the robustness of the {\method}, given the wide variety of challenging scenarios considered.}

\begin{table}[!t]
\caption{\small Evaluation with metrics in the scenarios considered for CALT*+M1 and the {\method} considering 9-11, 10, 11, 11-13 and 13 neighbours with $LoS=1$ relative to A*: T is the ratio of the computation time \revJAC{(related to efficiency)}, L is the ratio of the \revJAC{path} length, N is the ratio of the explored nodes, MD is the ratio of the mean distance to the close\revJAC{st} obstacles \revJAC{(related to safety)} and MA is the mean angle \revJAC{(related to \revJACS{geometric} smoothness)}.}
\centering
\scriptsize
\begin{tabular}{|p{0.01cm} c|c|c|c|c|c|c|c|c|c|}
\cline{3-7}\multicolumn{2}{c|}{} &CALT*+M1 & \method 9-11 & \method 10  & \method11  & \method 11-13\\
\hline
\multirow{5}{*}{S1} &
T $\downarrow$ & $3.653\pm0.073$ & $ \textbf{1.666$\pm$0.053} $ & $2.882\pm0.046$ & $ 2.297\pm0.062 $ & $ 2.811\pm0.052 $\\
& L $\downarrow$ & $ \textbf{0.966$\pm$0.004} $ & $ 1.020\pm0.022 $ & $1.011\pm0.009$ &  $1.001\pm0.013$ &  $0.990\pm0.011$\\
& N $\downarrow$ & $0.988\pm0.001$ & $ \textbf{0.541$\pm$0.009} $ & $0.9824\pm0.005$ & $ 0.733\pm0.014 $ & $ 0.867\pm0.004 $\\
& MD $\uparrow$ & $0.958\pm0.017$ & $ 0.884\pm0.115 $ & $ \textbf{1.158$\pm$0.119} $ & $ 0.992\pm0.071 $ & $ 0.943\pm0.056 $\\
& MA $\downarrow$ & $0.471\pm0.030$ & $ 0.532\pm0.046 $ & $0.467\pm0.022$ & $ 0.469\pm0.049 $ & $ \textbf{0.443$\pm$0.035} $\\
\hline
\multirow{5}{*}{S2 } &
T $\downarrow$ & $3.461\pm0.020$ & $ \textbf{1.827$\pm$0.017} $ & $2.830\pm0.055$ & $ 2.178\pm0.021 $ & $ 2.660\pm0.025 $\\
& L $\downarrow$ & $ \textbf{0.974$\pm$0.002} $ & $ 1.033\pm0.006 $ & $0.987\pm0.006$ &  $1.019\pm0.005$ &  $1.005\pm0.005$\\
& N $\downarrow$ & $0.998\pm0.001$ & $ \textbf{0.634$\pm$0.003} $ & $0.981\pm0.001$ & $ 0.761\pm0.002 $ & $ 0.877\pm0.004 $\\
& MD $\uparrow$ & $1.020\pm0.038$ & $ \textbf{1.050$\pm$0.061} $ & $0.951\pm0.034$ & $ 0.978\pm0.049 $ & $ 1.008\pm0.058 $\\
& MA $\downarrow$ & $0.524\pm0.014$ & $ 0.576\pm0.025 $ & $0.508\pm0.011$ & $ \textbf{0.540$\pm$0.016} $ & $ 0.459\pm0.013 $\\
\hline
\multirow{5}{*}{ S3} &
T $\downarrow$ & $3.398\pm0.041$ & $ \textbf{1.819$\pm$0.018} $ & $2.660\pm0.0174$ & $ 2.152\pm0.016 $ & $ 2.619\pm0.022 $\\
& L $\downarrow$ & $ \textbf{0.917$\pm$0.001} $ & $ 1.351\pm0.012 $ & $0.980\pm0.005$ &  $1.075\pm0.008$ &  $1.036\pm0.007$\\
& N $\downarrow$ & $0.998\pm0.001$ & $ \textbf{0.642$\pm$0.004} $ & $0.977\pm0.002$ & $ 0.779\pm0.005 $ & $ 0.874\pm0.005 $\\
& MD $\uparrow$ & $0.971\pm0.031$ & $ 0.678\pm0.017 $ & $ \textbf{1.042$\pm$0.030} $ & $ 0.796\pm0.021 $ & $ 0.934\pm0.047 $\\
& MA $\downarrow$ & $ \textbf{0.530$\pm$0.007} $ & $ 1.661\pm0.046 $ & $0.548\pm0.021$ & $ 1.099\pm0.048 $ & $ 0.638\pm0.031 $\\
\hline
\multirow{5}{*}{S4} &
T $\downarrow$ & $3.566\pm0.020$ & $ \textbf{1.328$\pm$0.018} $ & $2.732\pm0.019$ & $ 2.100\pm0.017 $ & $ 2.547\pm0.016 $\\
& L $\downarrow$ & $ \textbf{0.988$\pm$0.004} $ & $ 0.998\pm0.006 $ & $0.989\pm0.005$ &  $0.996\pm0.003$ &  $0.995\pm0.002$\\
& N $\downarrow$ & $1.004\pm0.001$ & $ \textbf{0.463$\pm$0.004} $ & $0.969\pm0.002$ & $ 0.738\pm0.003 $ & $ 0.833\pm0.004 $\\
& MD $\uparrow$ & $ \textbf{0.991$\pm$0.004} $ & $ 0.974\pm0.010 $ & $0.972\pm0.010$ & $ 0.969\pm0.004 $ & $ 0.989\pm0.003 $\\
& MA $\downarrow$ & $0.588\pm0.015$ & $ \textbf{0.537$\pm$0.007} $ & $0.545\pm0.018$ & $ 0.566\pm0.011 $ & $ 0.556\pm0.010 $\\
\hline
\multirow{5}{*}{ S5} &
T $\downarrow$ & $2.577\pm0.033$ & $ \textbf{0.782$\pm$0.034} $ & $1.791\pm0.052$ & $ 0.790\pm0.034 $ & $ 1.699\pm0.028 $\\
& L $\downarrow$ & $0.969\pm0.003$ & $ 0.967\pm0.009 $ & $0.962\pm0.005$ &  $ \textbf{0.952$\pm$0.004} $ &  $0.968\pm0.004$\\
& N $\downarrow$ & $0.997\pm0.004$ & $ \textbf{0.385$\pm$0.014} $ & $0.952\pm0.005$ & $ 0.406\pm0.013 $ & $ 0.760\pm0.004 $\\
& MD $\uparrow$ & $0.989\pm0.006$ & $ 0.993\pm0.014 $ & $0.964\pm0.005$ & $ 0.973\pm0.007 $ & $ \textbf{0.995$\pm$0.007} $\\
& MA $\downarrow$ & $0.430\pm0.013$ & $ 0.445\pm0.012 $ & $0.417\pm0.009$ & $ 0.414\pm0.018 $ & $ \textbf{0.413$\pm$0.016} $\\
\hline
\multirow{5}{*}{ S6} &
T $\downarrow$ & $2.807\pm0.029$ & $ \textbf{0.789$\pm$0.063} $ & $2.076\pm0.060$ & $ 0.862\pm0.119 $ & $ 1.722\pm0.037 $\\
& L $\downarrow$ & $ \textbf{0.972$\pm$0.043} $ & $ 1.062\pm0.027 $ & $0.996\pm0.015$ &  $1.028\pm0.036$ &  $0.976\pm0.011$\\
& N $\downarrow$ & $0.999\pm0.003$ & $ \textbf{0.387$\pm$0.033} $ & $0.974\pm0.026$ & $ 0.427\pm0.066 $ & $ 0.727\pm0.012 $\\
& MD $\uparrow$ & $0.987\pm0.007$ & $ 0.926\pm0.021 $ & $ \textbf{1.018$\pm$0.019} $ & $ 0.948\pm0.013 $ & $ 0.973\pm0.012 $\\
& MA $\downarrow$ & $0.670\pm0.052$ & $ 0.730\pm0.063 $ & $0.588\pm0.037$ & $ 0.624\pm0.044 $ & $ \textbf{0.571$\pm$0.033} $\\
\hline
\multirow{5}{*}{ S7} &
T $\downarrow$ & $2.227\pm0.020$ & $ \textbf{1.203$\pm$0.033} $ & $1.719\pm0.034$ & $ 1.365\pm0.031 $ & $ 1.613\pm0.014 $\\
& L $\downarrow$ & $ \textbf{0.983$\pm$0.002    } $ & $ 0.991\pm0.004 $ & $0.985\pm0.005$ &  $0.992\pm0.059$ &  $0.978\pm0.004$\\
& N $\downarrow$ & $0.999\pm0.002$ & $ \textbf{0.724$\pm$0.003} $ & $0.971\pm0.006$ & $ 0.789\pm0.005 $ & $ 0.873\pm0.002 $\\
& MD $\uparrow$ & $1.005\pm0.005$ & $ 1.005\pm0.007 $ & $ 0.997\pm0.010 $ & $ \textbf{1.011$\pm$0.010} $ & $ 0.987\pm0.009 $\\
& MA $\downarrow$ & $0.478\pm0.024$ & $ 0.511\pm0.028 $ & $0.490\pm0.022$ & $ 0.470\pm0.017 $ & $ \textbf{0.458$\pm$0.017} $\\
\hline
\multirow{5}{*}{ S8} &
T $\downarrow$ & $2.044\pm0.031$ & $ \textbf{1.038$\pm$0.032} $ & $1.604\pm0.032$ & $ 1.172\pm0.044 $ & $ 1.377\pm0.038 $\\
& L $\downarrow$ & $ \textbf{0.970$\pm$0.002} $ & $ 1.087\pm0.011 $ & $0.976\pm0.004$ &  $1.093\pm0.004$ &  $0.973\pm0.001$\\
& N $\downarrow$ & $0.994\pm0.001$ & $ \textbf{0.686$\pm$0.017} $ & $1.014\pm0.016$ & $ 0.734\pm0.023 $ & $ 0.831\pm0.009 $\\
& MD $\uparrow$ & $\textbf{1.004$\pm$0.019}$ & $ 0.909\pm0.021 $ & $ 0.988\pm0.025 $ & $ 0.977\pm0.011 $ & $ 0.978\pm0.004 $\\
& MA $\downarrow$ & $0.405\pm0.021$ & $ 0.370\pm0.025 $ & $0.438\pm0.021$ & $ \textbf{0.363$\pm$0.010} $ & $ 0.419\pm0.014 $\\
\hline
\end{tabular}
\label{table_results}
\end{table}

\section{Conclusion and Future Work} \label{sec:conclusion}

A new heuristic search planner based on a modified Lazy Theta* planning algorithm that integrates EDF and exploits its properties has been presented. We proposed the FS-Planner\revJAC{, which} generates fast and safe paths and eliminates the need for post-processing methods to smooth the paths. A new cost function has been \revJAC{introduced} that \revJAC{incorporates} a term related to the distance from obstacles\revJAC{, thus providing} a safer path planner. 

The exploitation of the analytical properties of EDFs have enabled faster computation of the EDF\revJAC{-based} cost along segments. We demonstrate the proposed approximation of $O(s_{i},s'_{i+1})$ based on the analytical properties of the EDF (\ref{eq:approx}).
Thanks to the EDF properties\revJAC{, the number of visibility neighbours can also be reduced} during path exploration \revJAC{using} the obstacle-gradient information. Moreover, the way in which visibility neighbours \revJAC{are selected directly} influences the safety of the path.

Heuristic search planners as Lazy Theta* algorithm take advantage of the triangle inequality to simplify the calculations to select the parent node. We have demonstrated that the EDF-based cost function used in the {\method} satisfies the triangle inequality and can be implemented in any heuristic algorithm such as A* or Theta*. Therefore, the {\method} improves \revJAC{upon} CALT*+M1 because it does hold the triangle inequality, resulting in fewer calculations.



These claims have been validated with \revJAC{extensive} tests and comparisons in 3D challenging scenarios and setups. \revJACS{In addition to simulated indoor environments, we also evaluated the {\method} in three real-world outdoor scenarios, which provide complex and irregular obstacle configurations. These real-environment experiments complement the simulation study and further support the robustness and practical applicability of the {\method}.}


\revJAC{Although the EDF-based cost term has been implemented here within a heuristic graph-search planner, the formulation itself is orthogonal to the underlying planning algorithm. Because the cost satisfies the triangle inequality and leverages analytical EDF properties, it can be incorporated into other planning paradigms, such as sampling-based methods, to potentially improve their safety and their ability to reason about obstacle proximity.}

\revJACS{Overall, the {\method} deliberately trades strict theoretical guarantees of completeness and optimality for significant practical gains in computational efficiency. By reducing the number of explored visibility neighbours, the planner becomes sub-optimal and potentially incomplete; however, this relaxation leads to a structural reduction of the search space, which is particularly impactful in 3D environments with high branching factors. Although the fallback mechanism restores completeness by reverting to full neighbour exploration when required, it was never activated in any of the evaluated scenarios, including complex 3D and real-world environments. This observation should be interpreted as empirical evidence of robust behaviour rather than as a formal guarantee.}

\revJACS{Furthermore, the benefits of the {\method} are not limited to computation time alone. The experimental results show that the reduction in computational cost, approximately $2\times$ – $3\times$, is accompanied by consistent improvements in geometric smoothness of the path, reflected in smaller heading changes, and by equal or higher safety levels in terms of obstacle clearance. Moreover, the {\method} offers an adjustable trade-off between efficiency and guarantees: configurations with fewer neighbours prioritise speed, while configurations with more neighbours approach optimal solutions. This flexibility makes the {\method} particularly suitable for local planning scenarios, where real-time performance, safety, and smoothnees are often more critical than strict optimality.}

Future work will \revJAC{also include experiments with aerial robots, as well as} the integration of new methods for online EDF computation based on neural radiation fields (NeRF). \revJAC{In addition,} we will analyze how local path planners can take advantage of such online EDFs.

\section*{Acknowledgements}
This work was partially supported by the grants INSERTION (PID2021-127648OB-C31) and NORDIC (TED2021-132476B-I00), both funded by the ``Agencia Estatal de Investigación -- Ministerio de Ciencia, Innovación y Universidades'' and the ``European Union NextGenerationEU/PRTR''.


\bibliographystyle{elsarticle-num}
\bibliography{bibliography_ras}






\end{document}